\documentclass{article}

\usepackage{arxiv}

\usepackage[utf8]{inputenc} 
\usepackage[T1]{fontenc}    
\usepackage{hyperref}       
\usepackage{url}            
\usepackage{booktabs}       
\usepackage{amsfonts}       
\usepackage{nicefrac}       
\usepackage{microtype}      
\usepackage{lipsum}		
\usepackage{graphicx}
\usepackage{natbib}
\usepackage{doi}
\usepackage{physics}
\usepackage{float}
\usepackage{subcaption}
\usepackage{multirow}
\usepackage{xcolor}
\usepackage{romannum}

\usepackage[authormarkup=none]{changes}
\definechangesauthor[name={Daniel Badawi}, color=red]{ec}

\title{Physics-Informed Neural Network for the Transient Diffusivity Equation in Reservoir Engineering}

\author{
	Daniel Badawi\\
	Department of Petroleum Engineering\\
	Texas A\&M University\\
	daniel\_88@tamu.edu\\
	\\
	\And
	Eduardo Gildin\\
	Department of Petroleum Engineering\\
	Texas A\&M University\\
	egildin@tamu.edu\\
	\\
}

\date{}

\renewcommand{\shorttitle}{PINN for the Transient Diffusivity Equation in Reservoir Engineering}

\hypersetup{
    pdftitle={Physics-Informed Neural Network for the Transient Diffusivity Equation in Reservoir Engineering},
    pdfsubject={},
    pdfauthor={Daniel Badawi, Eduardo Gildin},
    pdfkeywords={Physics-informed Neural Networks (PINNs), Physics-informed Machine Learning, Diffusivity Equation, Reservoir Engineering, Buildup Tests},
}

\begin{document}
\pagenumbering{arabic}
\maketitle

\begin{abstract}
	Physics-Informed machine learning models have recently emerged with some interesting and unique features that can be applied to reservoir engineering. In particular, physics-informed neural networks (PINN) leverage the fact that neural networks are a type of universal function approximators that can embed the knowledge of any physical laws that govern a given data-set in the learning process, and can be described by partial differential equations. The transient diffusivity equation is a fundamental equation in reservoir engineering and the general solution to this equation forms the basis for Pressure Transient Analysis (PTA). The diffusivity equation is derived  by combining three physical principles, the continuity  equation, Darcy's equation, and the equation of state for a slightly compressible liquid. Obtaining general solutions to this equation is imperative to understand flow regimes in porous media. Analytical solutions of the transient diffusivity equation are usually hard to obtain due to the stiff nature of the equation caused by the steep gradients of the pressure near the well. In this work we apply physics-informed neural networks to the one and two dimensional diffusivity equation and demonstrate that decomposing the space domain into very few subdomains can overcome the stiffness problem of the equation. Additionally, we demonstrate that the inverse capabilities of PINNs can estimate missing physics such as permeability and distance from sealing boundary similar to buildup tests without shutting in the well.
\end{abstract}


\section{Introduction}
\label{sec:intro}

The diffusivity equation is a stiff spatiotemporal non-linear partial differential equation (PDE) that describes fluids flow in porous media. It results from combining three physical principles, the material balance equation, Darcy's law and the equation of state of slightly compressible liquid. Thus it is regarded the most important equation in reservoir engineering. Analytical solutions for some simplified forms of the diffusivity equation exist however their practical applications are very limited especially for history matching applications. For single-phase infinite-acting circular reservoirs analytical solution is obtained using Boltzmann's variable transformation which transforms the PDE to an ordinary differential equation (ODE). For circular bounded reservoirs analytical solutions are available in Laplace domain and thus require inversion algorithms \citep{stehfest_algorithm, Talbot_algorithm, dehoog_knight_stokes} for the PDE solution to be expressed in the temporal domain.

The stiffness of the PDE which is caused by the steep gradients near the well poses a major challenge even to numerical methods where very fine meshes need to be considered. Numerical methods are very good tools to solve the diffusivity equation, in fact it is the essence of reservoir simulations. However for well testing techniques which are inverse problems, numerical methods are very limited. Another drawback of numerical methods is that they only offer solutions at discrete specified times which limits our ability to calculate the flow rates at all times. For that, we prefer to work with continuous analytical solutions or any good approximation of analytical solutions.

Recently, there is noticeable and growing increase in the use of machine learning (ML) applications in various areas in science and engineering which most of them are fully data-driven. One major shortcoming of data-driven machine learning is the failure to honor any physical relationship between inputs and outputs. Consequently, they suffer to extrapolate predictions outside the training domain and therefore falls short when applied to time-dependent reservoir engineering applications where temporal extrapolation is essential.

A PINN is a scientific machine learning technique used to solve problems involving PDEs \citep{raissi}. PINNs approximate PDE solutions by training a neural network to minimize a loss function; it includes terms reflecting the initial and boundary conditions along the space-time domain’s boundary and the PDE residual at selected points in the domain (called collocation points). Incorporating a residual network that encodes the governing physics equations is a significant novelty with PINNs. The basic concept behind PINN training is that it can be thought of as an unsupervised strategy that does not require labelled data, such as results from prior simulations or experiments. The PINN algorithm is essentially a mesh-free technique that finds PDE solutions by converting the problem of directly solving the governing equations to a loss function optimization problem. It works by integrating the mathematical model into the network and reinforcing the loss function with a residual term from the governing equation \citep{whatsnext}. Additionally, using neural network back-propagation, when a PINN model is trained, not only the PDE solution is obtained but also the solution derivatives w.r.t inputs are also obtained. Therefore, PINNs can be regarded as alternatives to analytical solutions.

In this paper, our work can be broadly split into two sections. In section 1, we apply the PINN methodology to the 1D cartesian diffusivity equation and show how to utilize the inverse capabilities of PINNs to predict some unknown reservoir properties which previously were estimated using well testing techniques. In section 2, we apply PINNs to the radial diffusivity equation in a circular bounded reservoir under both constant bottom-hole pressure (BHP) and under constant production flow rate. We also present the domain-decomposition method to tackle the stiffness of the radial diffusivity equation.

\section{Background}
\label{sec:background}

\subsection{Diffusivity Equation}
\begin{figure}[ht]
    \centering
    \includegraphics[height=189px, width=268px]{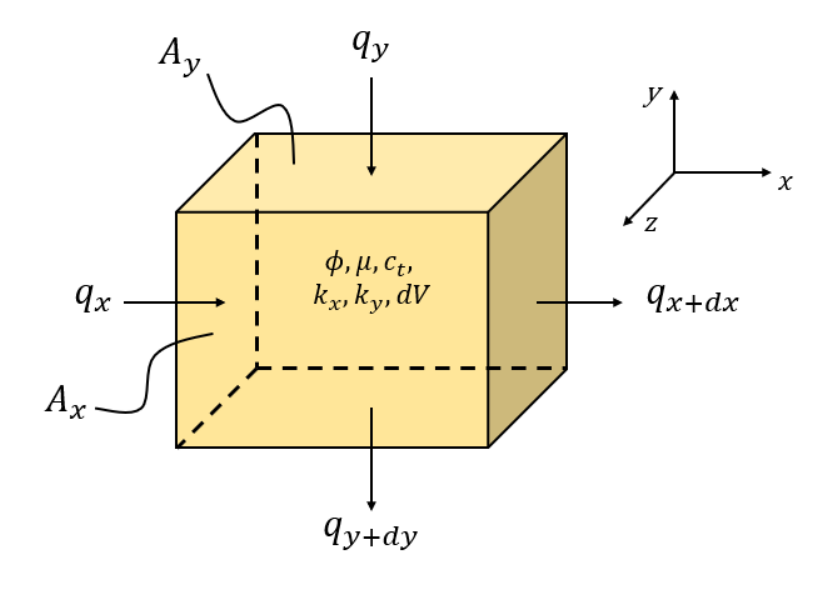}
    \caption{Infinitesimal Cartesian volume element for the diffusivity equation analysis.}
    \label{fig:box_img}
\end{figure}

The diffusivity equation describes the flow in porous media and can be derived using the material balance equation and Darcy's law. Taking an infinitesimal volume element as in Fig. \ref{fig:box_img}, the material balance states that the difference between the mass flow rate entering the volume element and the mass flow rate leaving the volume element must equal the rate of change of mass accumulating in the volume element due expansion of the pore volume and change of mass contributed by sources/sinks. Assuming 2D case and putting in equation form gives the following:

\begin{align}\label{material_balance_eq_1}
    \underbrace{\big(q \rho \big\rvert_{x} + q \rho \big\rvert_{y}\big)}_{\substack{\text{Mass flow rate} \\\\ \text{IN}}} - \underbrace{\big(q \rho \big\rvert_{x+dx} + q \rho \big\rvert_{y+dy}\big)}_{\substack{\text{Mass flow rate} \\\\ \text{OUT}}} = \underbrace{\hspace{0.5cm}\frac{\partial m}{\partial t}\hspace{0.5cm}}_{\substack{\text{Rate of change } \\ \text{of mass in the} \\ \text{ volume element}}} + \underbrace{\hspace{0.3cm} q_{ss} \cdot dV\hspace{0.3cm}}_{\substack{\text{sources / sinks}}}
\end{align}

where $q\rho \big\rvert_{x}$ is the flux at $x$. $q$ is the volumetric flux in $[m^3/sec]$, $\rho$ is the density of the fluid in $[kg/m^3]$ and $q_{ss}$ is the sources/sinks term in $[kg/m^3]$. In this work we don't have sources or sinks and therefore the term will be neglected. $m$ is the accumulated mass inside the element volume and is given by:
\begin{align*}
    m=\rho dV
\end{align*}
with $dV$ is the volume element in $[m^3]$. It is important to note that the fluid can occupy a fraction of the element volume called the pore volume and is given by:
\begin{align*}
\begin{split}
    dV_p &= \phi \cdot dV \\
    dV &= dx\cdot dy \cdot dz
\end{split}  
\end{align*}
where $\phi$ is the porosity of the element volume $dV$ and defined as the volume that can be occupied by the fluid divided by the total geometric volume. As a result of the definition $\phi\in [0,1]$. Therefore, the element volume $dV$ can be divided into two volumes: pore volume $dV_p$ and rock volume $dV_r$.
\begin{align*}
    dV = dV_p+dV_r
\end{align*}
\begin{align}\label{material_balance_eq_2}
    -\big(q \rho \big\rvert_{x+dx} - q \rho \big\rvert_{x}\big) - \big(q \rho \big\rvert_{y+dy} - q \rho \big\rvert_{y}\big)  = dV\frac{\partial \big(\phi \rho\big)}{\partial t}
\end{align}

\begin{align}\label{material_balance_eq_3}
    -\frac{1}{A_x}\frac{\big(q \rho \big\rvert_{x+dx} - q \rho \big\rvert_{x}\big)}{dx} -\frac{1}{A_y}\frac{\big(q \rho \big\rvert_{y+dy} - q \rho \big\rvert_{y}\big)}{dy}  = \frac{\partial \big(\phi \rho\big)}{\partial t}
\end{align}

where $A_x = dy \cdot dz$ and $A_y = dx \cdot dz$.

Simplifying eq. \ref{material_balance_eq_3} gives:
\begin{align}\label{material_balance_eq_4}
    -\frac{1}{A_x}\frac{\partial \big(q \rho\big)_x}{\partial x} -\frac{1}{A_y}\frac{\partial \big(q \rho\big)_y}{\partial y}  = \frac{\partial \big(\phi \rho\big)}{\partial t}
\end{align}
\\
Darcy's law \citep{darcy} can be mathematically written as the following:
\begin{align}\label{darcy_law}
    q=-\frac{kA}{\mu} (\nabla p+\rho g \nabla z)
\end{align}
\\
where $q$ is the volumetric flux in $[m^3/sec]$, $k$ is the permeability $[m^2]$, $A$ is the flux area in $[m^2]$, $\mu$ is the fluid viscosity in [$pa\cdot sec$], $\nabla p$ is the pressure gradient, $\rho$ is the fluid density in $[kg/m^3]$, $g$ is the gravity constant in $[m/sec^2]$, and $\nabla z$ is the gradient of the elevation ($z$ is in the gravity direction). Neglecting the gravity term, the 2D Darcy's law can be expressed as:
\begin{subequations}
\begin{align}
 q_x=-\frac{k_xA_x}{\mu} \frac{\partial p}{\partial x}\label{darcy_x}\\
 q_y=-\frac{k_yA_y}{\mu} \frac{\partial p}{\partial y}\label{darcy_y}
\end{align}
\end{subequations}
where $A_x$ and $A_y$ is the flux area perpendicular to $x$ direction and $y$ direction respectively. $k_x$ and $k_y$ is the permeability in $x$ and $y$ direction respectively.\\
Substituting eqs. \ref{darcy_x}, \ref{darcy_y} into eq. \ref{material_balance_eq_4} and expanding the right hand side gives:\\
\begin{align}\label{diff_eq_1}
    \frac{\partial}{\partial x} \bigg(\rho \frac{k_x}{\mu} \frac{\partial p}{\partial x}\bigg) + \frac{\partial}{\partial y} \bigg(\rho \frac{k_y}{\mu} \frac{\partial p}{\partial y}\bigg)  = \phi\frac{\partial \rho}{\partial t} + \rho\frac{\partial \phi}{\partial t}
\end{align}
The time derivative of the density appearing on the right of eq. \ref{diff_eq_1} can be expressed in terms of a time derivative of the pressure by using the basic thermo-dynamic definition of isothermal compressibility of fluids:
\begin{align}\label{fluid_compressibility_1}
    c_{f} = - \frac{1}{V_f}\frac{\partial V_f}{\partial p}
\end{align}
where $c_f$ is the fluid compressibility in $[1/pa]$, $V_f$ is the fluid volume in $[m^3]$ and $p$ is the pressure acting on the fluid in the pore volume (pore pressure) in $[pa]$, and since
\begin{align*}
    \rho=\frac{m}{V_f}
\end{align*}
then the compressibility can be alternatively expressed as:
\begin{align}\label{fluid_compressibility_2}
    c_f=-\frac{\rho}{m}\frac{\partial}{\partial p} \bigg(\frac{m}{\rho}\bigg)=\frac{1}{\rho}\frac{\partial \rho}{\partial p}
\end{align}
and differentiating with respect to time gives:
\begin{align}\label{drho_dt}
    \rho c_f \frac{\partial p}{\partial t} = \frac{\partial \rho}{\partial t}
\end{align}
The pore volume also changes with pressure, therefore similar to the fluid compressibility the rock compressibility is defined as:
\begin{align}\label{rock_compressibility_1}
    c_r = \frac{1}{V_p}\frac{\partial V_p}{\partial p}
\end{align}
Notice that eq. \ref{rock_compressibility_1} has no negative sign, this because the rock compressibility is expressed in terms of pore volume $V_p$ and not rock volume $V_r$, and since
\begin{align*}
    V_p = \phi V
\end{align*}
eq. \ref{rock_compressibility_1} can be expressed as:
\begin{align}\label{rock_compressibility_2}
    c_r = \frac{1}{\phi}\frac{\partial \phi}{\partial p}
\end{align}
and differentiating with respect to time gives:
\begin{align}\label{dphi_dt}
    \phi c_r \frac{\partial p}{\partial t} = \frac{\partial \phi}{\partial t}
\end{align}
\\
finally, substituting eqs. \ref{drho_dt} and \ref{dphi_dt} in eq. \ref{diff_eq_1} reduces to the latter:\newline
\begin{align}\label{diff_eq_2}
    \frac{\partial}{\partial x} \bigg(\rho \frac{k_x}{\mu} \frac{\partial p}{\partial x}\bigg) + \frac{\partial}{\partial y} \bigg(\rho \frac{k_y}{\mu} \frac{\partial p}{\partial y}\bigg)  = \phi c_t \rho \frac{\partial p}{\partial t}
\end{align}\newline[2pt]
where $c_t$ is the total compressibility and is given by:
\begin{align*}
    c_t = c_f + c_r
\end{align*}
Eq. \ref{diff_eq_2} is the basic 2D, partial differential equation of any single phase fluid in a porous medium. The equation of course is highly non-linear due to the implicit pressure dependence of the density, viscosity, and compressibility. For that, it is almost hopeless to find simple analytical solutions of the equation. One option to reduce the non-linearity of the equation is to assume constant compressibility, viscosity, and density and this yields to the following:
\begin{align}\label{diff_eq_3}
\begin{split}
    \frac{\partial}{\partial x} \bigg(k_x \frac{\partial p}{\partial x}\bigg) + \frac{\partial}{\partial y} \bigg( k_y \frac{\partial p}{\partial y}\bigg)  = \phi \mu c_t \frac{\partial p}{\partial t}\\[5pt]
    x \in [x_{e1}, x_{e2}], \hspace{0.3cm} y\in[y_{e1}, y_{e2}], \hspace{0.3cm} t\in[0, t_f]
\end{split}
\end{align}

In this paper we will work with the 1D Cartesian case, thus eq. \ref{diff_eq_3} reduces to:
\begin{align}\label{1d_diff_eq}
\begin{split}
    \frac{\partial}{\partial x} \bigg(k_x \frac{\partial p}{\partial x}\bigg)= \phi \mu c_t \frac{\partial p}{\partial t}\\[5pt]
    x \in [x_{e1}, x_{e2}], \hspace{0.3cm} t\in[0, t_f]
\end{split}
\end{align}
Eq. \ref{diff_eq_3} can be expressed in cylindrical coordinates as follows:
\begin{align}\label{diff_eq_cylindrical}
\begin{split}
    \frac{1}{r}\frac{\partial}{\partial r}\bigg(k_r r\frac{\partial p}{\partial r}\bigg) = \phi \mu c_t \frac{\partial p}{\partial t}\\[5pt]
    r \in [r_w, r_e], \hspace{0.3cm} t\in[0, t_f]
\end{split}
\end{align}

where $k_r$ is the permeability in the radial direction in $[m^2]$, $r_w$ and $r_e$ are the wellbore radius and the outer reservoir radius in $[m]$ respectively. As mentioned above, eq. \ref{diff_eq_cylindrical} is a stiff PDE and that is due to the steep pressure gradients near the well. One reason for the steep gradients is the small wellbore radius $r_w$ which usually ranges from $3[in]$ to $5[in]$. Equations \ref{1d_diff_eq} \ref{diff_eq_cylindrical} are two-point boundary value problems that require an initial condition and two boundary conditions.

\section{Buildup Test}
\label{sec:builduptest}

\begin{figure}[H]
    \includegraphics[height=208px, width=379px]{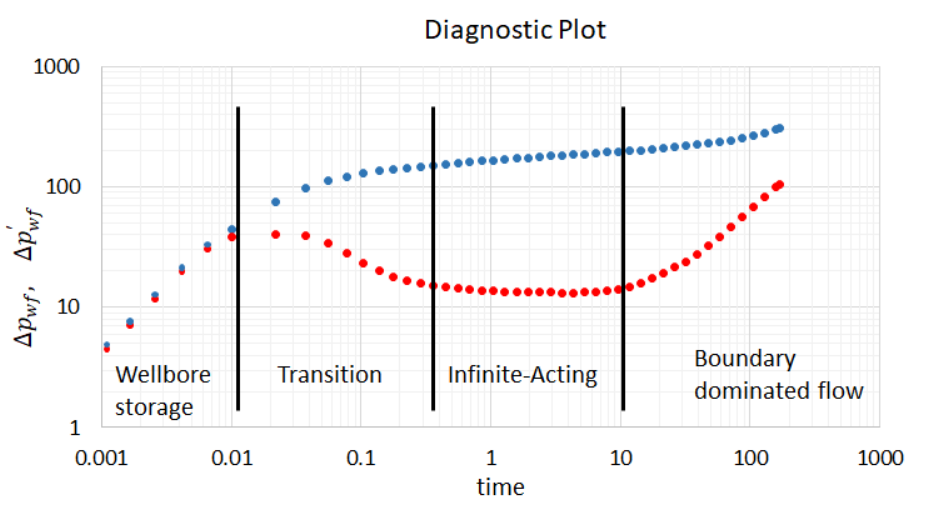}
    \centering
    \caption{Diagnostic plot. The blue points are the bottom-hole pressure recording after well shut-in. The red points are the derivative of the recorded bottom-hole pressure with respect to shut-in time.}
    \label{fig:diagnostic_plot}
\end{figure}

Buildup tests are perhaps the most popular tests in well testing. They are utilized to estimate reservoir properties such as average reservoir permeability, distance from boundary (fault), wellbore storage, skin factor, and many more. A Traditional build up test is performed to a well producing under constant flow rate. The test starts by shutting in the well and recording the bottom-hole pressure change as a function of time. Then the data is plotted on a diagnostic plot. A diagnostic plot is a log-log plot of the pressure change and pressure derivative (vertical axis) versus elapsed time (horizontal axis). It is qualitative plot used to identify flow regimes at different time periods. For each flow regime period we fit an analytical solution and from that we can obtain the reservoir properties. A typical diagnostic plot looks similar to Fig. \ref{fig:diagnostic_plot}. At early times the pressure buildup is dominated by wellbore storage, this is when $\Delta p_{wf}$ and $\Delta p'_{wf}$ are equal with slopes equal to unity, next we have a transition period, followed by infinite acting period, this is when $\Delta p'_{wf}$ is constant, and finally boundary dominated period this is when the boundaries felt the shut-in.

As mentioned previously, under infinite-acting conditions and constant permeability $k_r$, eq. \ref{diff_eq_cylindrical} has analytical solution in terms of exponential integral
\begin{align}\label{analytical_solution_1}
    p_{wf}&=p_i-\frac{q\mu}{2\pi k_rh}ei\bigg(\frac{\phi \mu c_t r_w^2}{ 4k_rt}\bigg)
\end{align}
and since
\begin{align*}
    ei(x)\approx-ln(\gamma x) \hspace{1cm} \text{for} \hspace{0.1cm} x<0.01
\end{align*}
where $\gamma = 1.781$. Therefore, eq. \ref{analytical_solution_1} can be approximated as\newline
\begin{align}\label{analytical_solution_2}
    p_{wf}&=p_i-\frac{q\mu}{2\pi k_rh}ln\bigg(\frac{4k_rt}{\gamma \phi \mu c_t r_w^2}\bigg)\\[10pt]
    \Delta p_{wf}  &\equiv p_i - p_{wf}=\frac{q\mu}{2\pi k_rh}ln\bigg(\frac{4k_rt}{\gamma \phi \mu c_t r_w^2}\bigg)\label{analytical_solution_3}
\end{align}
where $p_i$ is the reservoir initial pressure, $p_{wf}$ is the bottom-hole pressure, $q$ is the flow rate prior to shut-in. $\gamma = 1.781$, $t$ is time, and $h$ is the producing layer thickness in $[m]$.\\\\
Diagnostic plots are qualitative plots used to identify different flow periods. For quantitative analysis specialized plots are used. Each flow period has its own specialized plot, for example, the specialized plot associated with the infinite-acting period is a semi-log plot with pressure change and its derivative on the linear vertical axis and elapsed time on the log horizontal axis. The reason for this is simply because it is evident from eq. \ref{analytical_solution_2} that the pressure change and elapsed time have a linear relationship on the semi-log plot. Therefore if we plot the pressure data on a semi-log specialized plot we can quantify the value $m$ which is the slope and subsequently we can estimate $k_r$ using eq. \ref{slope_2}.
\begin{align}
    m &= \frac{\partial \Delta p_{wf}}{\partial ln(t)} =\frac{q\mu}{2\pi k_rh}\label{slope_1}\\
    k_r&=\frac{q\mu}{2\pi mh}\label{slope_2}
\end{align} 

\subsection{Initial and Boundary Conditions}
\label{sec:ic_bc}
This work is divided into two sections. Section 1 is associated with eq. \ref{1d_diff_eq} and Section 2 is associated with eq. \ref{diff_eq_cylindrical} and both of these equations are two-point boundary problems thus, need an initial condition and two boundary conditions. Each section has two cases in which they differ by their boundary conditions. For all cases of both sections the initial condition is defined as:
\begin{align}\label{ic_eq}
      p(r,t=0) = p_0 = 25.0 \hspace{0.1cm}[MPa]\tag{2.20}
\end{align}
  
\textbf{Section 1:}\\
\begin{itemize}
      \item \textbf{Case 1.1}, the well is controlled by constant BHP at $x_w=0$, and a no-flow boundary at $x_e$, thus:
\begin{align*}
      p(x_{w},t) &= p_{wf} = 3.0 \hspace{0.1cm} [MPa]\\[10pt]
      x_e\frac{\partial p}{\partial x}\Big\rvert_{x_e} &= 0
\end{align*}
      
\item \textbf{Case 1.2}, the well is placed inside the domain $x_w \in [-500, 750]$ at $x_w=0$, and controlled by constant BHP. Also, two no-flow boundaries are set at $x_{e1}=-500$ and $x_{e2}=750[m]$, thus:
\begin{align*}
      p(x_w=0, t) = p_{wf} = 3.0 \hspace{0.1cm} [MPa]\\[10pt]
      x_{e1}\cdot \frac{\partial p}{\partial x}\Big\rvert_{x_{e1}} = 
      x_{e2}\cdot \frac{\partial p}{\partial x}\Big\rvert_{x_{e2}} = 0
\end{align*}
  \end{itemize}
  
  \begin{figure}[H]
     \includegraphics[height=114px, width=364px]{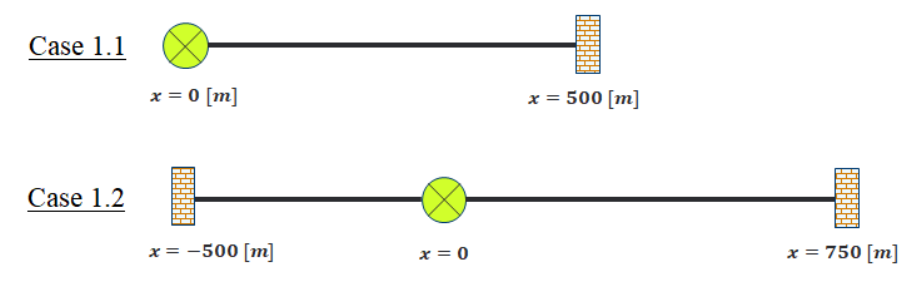}
     \centering
     \caption{Schematic representation of cases 1.1 and 1.2. The green circle is the well and the brick walls are the no-flow boundaries. }
     \label{fig:case2.1_and_case_2.2_img}
  \end{figure}

  \textbf{Section 2:}
  
  \begin{itemize}
      \item For both cases of section \Romannum{2}, the outer boundary is a         no-flow boundary, that is:
      \begin{align*}\label{noflow}
           q_e = 0
      \end{align*}
      Using Darcy's law:
      \begin{equation*} \label{darcy's}
         q(r, t) = \frac{kA}{\mu}\frac{\partial p}{\partial r}
      \end{equation*}
      $A$ is the flux area. The flux area at the boundary is given by:
      \begin{equation*}
          A_e = 2\pi r_e h
      \end{equation*}
      $h$ is the producing layer thickness. Combining the above yields to:
      \begin{align*}
           r_e\frac{\partial p}{\partial r}\Big\rvert_{r_e} = 0 
      \end{align*}
      \item \textbf{Case 2.1}, the well is producing at a constant BHP:
      \begin{equation*} \label{well_BC}
         p(r_{w},t) = p_{wf} = 3.0 \hspace{0.1cm} [MPa]
      \end{equation*}
      
      \item \textbf{Case 2.2}, the well is producing at a constant rate:
      \begin{equation*} \label{q_w}
         r_w\frac{\partial p}{\partial r}\Big\rvert_{r_w} = \frac{q_w\mu}{2\pi k h}=p_{ch}
      \end{equation*}
  \end{itemize}
  
  \begin{figure}[H]
     \includegraphics[height=117px, width=181px]{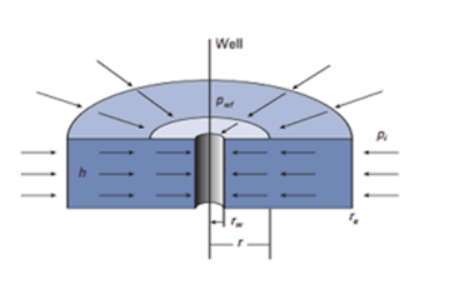}
     \centering
     \caption{Schematic of the 2D cases of Section \Romannum{2}.}
     \label{fig:2D_schematic}
  \end{figure}

\subsection{Physics-Informed Neural Networks}
\label{sec:pinns}
\begin{figure}[H]
     \includegraphics[height=234px, width=400px]{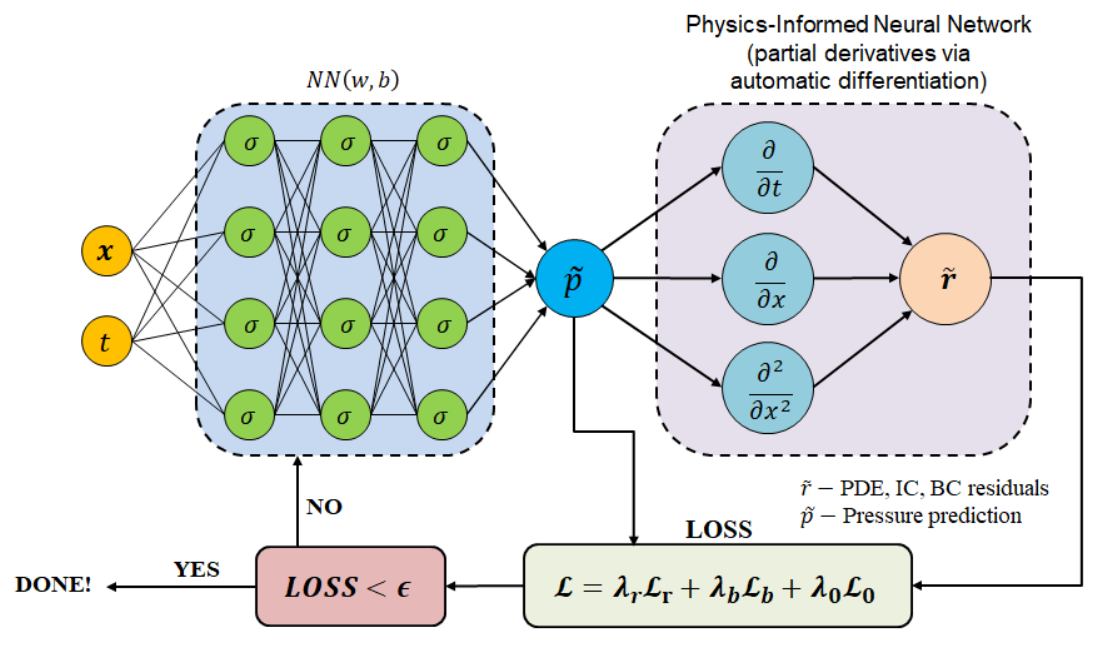}
     \centering
     \caption{Schematic of a physics-informed neural network.}
     \label{fig:pinn_architecture_img}
  \end{figure}

PINNs can be used to approximate solutions of PDEs. For a general PDE of the form: 
  \begin{subequations}
  \begin{align}
     &\frac{\partial u(\boldsymbol{x},t)}{\partial t} + \mathcal{N}_x(u(\boldsymbol{x},t))=0, \hspace{0.5cm} \boldsymbol{x}\in \Omega, t\in[0,T]\label{pinn_pde}\\
      &u(\boldsymbol{x},t) = g(\boldsymbol{x},t), \hspace{0.3cm} \boldsymbol{x}\in \partial \Omega, t\in[0,T]\label{pinn_bc}\\
      &u(\boldsymbol{x},0) = u_0(\boldsymbol{x}), \hspace{0.3cm} \boldsymbol{x}\in \Omega\label{pinn_ic}
  \end{align}
  \end{subequations}
  where $u(\boldsymbol{x}, t)$ is the solution to eq. \ref{pinn_pde}, $\Omega$ is the space domain, $x$ is a spatial vector variable, $t$ is time, and $\mathcal{N}_x(.)$ is a differential operator. Equations \ref{pinn_bc}, \ref{pinn_ic} are the boundary and initial conditions respectively. Physics-informed neural network $\Tilde{u}(\boldsymbol{x},t, \boldsymbol{w})$ can approximate the solution of eq. \ref{pinn_pde}, where $\boldsymbol{w}$ are the weights and biases of the PINN (\citep{raissi}). The PINN is trained by minimizing the loss function which is given as follows: 
  \begin{align}
      \mathcal{L}(\boldsymbol{w}) = \lambda_r\mathcal{L}_r(\boldsymbol{w}) + \lambda_b\mathcal{L}_b(\boldsymbol{w}) + \lambda_0\mathcal{L}_0(\boldsymbol{w})
  \end{align}
  
    where $\lambda_r, \lambda_b, \lambda_0\in \mathbb{R}$ are weights and $\mathcal{L}_r(w), \mathcal{L}_b(w), \mathcal{L}_0(w)$ correspond to the residual, boundary conditions, and initial conditions accordingly, and are given by:
  \begin{subequations}
  \begin{align}
      \mathcal{L}_r(\boldsymbol{w}) =& \frac{1}{N_r} \sum_{i=1}^{N_r}r({\boldsymbol{x}^i_r}, {t^i_r}, \boldsymbol{w}),\label{loss_res}\\
      \mathcal{L}_b(\boldsymbol{w}) =& \frac{1}{N_b} \sum_{i=1}^{N_b}\big(\Tilde{u}({\boldsymbol{x}^i_b}, {t^i_b}, \boldsymbol{w})-g(\boldsymbol{x}^i_b, t^i_b)\big)^2,\label{loss_bc}\\
      \mathcal{L}_0(\boldsymbol{w}) =& \frac{1}{N_0} \sum_{i=1}^{N_0}\big(\Tilde{u}({\boldsymbol{x}^i_0}, {0^i}, \boldsymbol{w})-u_0(\boldsymbol{x}^i_0)\big)^2,\label{loss_ic}
  \end{align}
  \end{subequations}
  where $\{x_r^i, t_r^i\}_{i=1}^{N_r}$ is a collocation points list sampled from the spatial and temporal domain \ref{pinn_pde}, $\{x_b^i, t_b^i\}_{i=1}^{N_b}$ is a boundary points list sampled from the boundaries \ref{pinn_bc}, and $\{x_0^i, 0^i\}_{i=1}^{N_0}$ is an initial condition points list sampled from the initial domain \ref{pinn_ic}. These points is where the loss function components $\mathcal{L}_r(w), \mathcal{L}_b(w), \mathcal{L}_0(w)$ are minimized respectively. $r(\boldsymbol{x},t;\boldsymbol{w})$ is the residual of the PDE and is defined as:
  \begin{equation}
      r(\boldsymbol{x},t;\boldsymbol{w}) = \frac{\partial \Tilde{u}(\boldsymbol{x},t;\boldsymbol{w})}{\partial t} + \mathcal{N}_x(\Tilde{u}(\boldsymbol{x},t, \boldsymbol{w}))
  \end{equation}
  The training of the PINN is achieved by gradient descent, where the gradients are calculated via automatic differentiation \citep{auto_diff}. The gradient descent minimization algorithm is expressed by eq.\ref{minimization_algorithm} where $\alpha$ is the learning rate. The architecture of a physics-informed neural network can be schematically visualized as in Fig. \ref{fig:pinn_architecture_img}.
  
  \begin{align}\label{minimization_algorithm}
      \boldsymbol{w}_{j+1} = \boldsymbol{w}_j - \alpha \frac{\partial \mathcal{L}}{\partial \boldsymbol{w}_j}
  \end{align}
  
\section{Physics-Informed Neural Network Setup}
\label{pinn_setup}

\subsection{Equation Scaling and Domain Generation}
In this section we show how to reform the diffusivity equation to fit the requirements of PINNs. In general, for efficient training of neural networks the inputs and outputs are scaled between $[0,1]$ or $[-1,1]$. In our study both inputs (space-time) and the output (pressure) are scaled between $[0,1]$ unless stated otherwise.

\begin{figure}[H]
\centering
\begin{subfigure}{0.45\textwidth}
        \includegraphics[height=123px, width=174px]{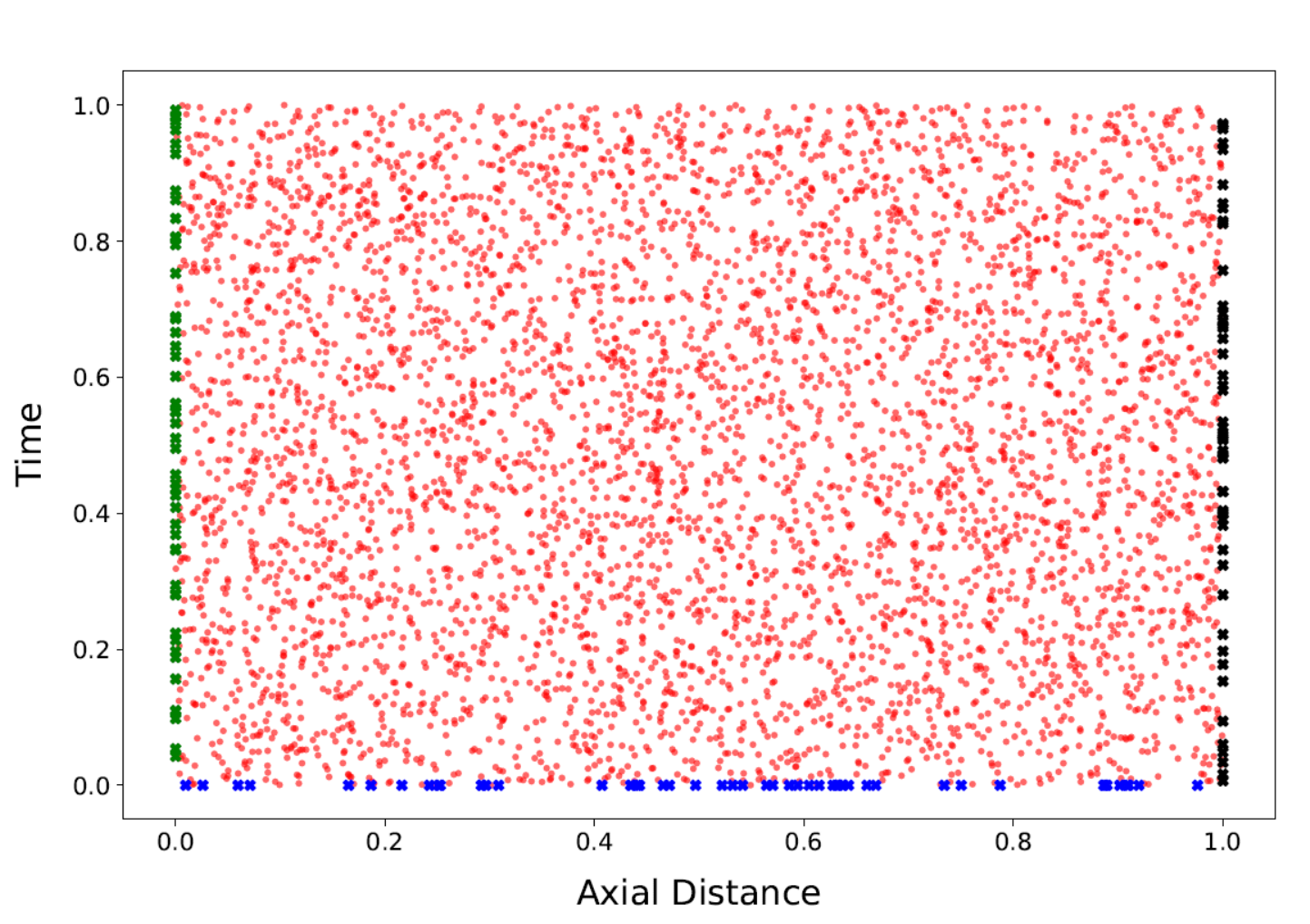}
    \caption{}
    \label{fig:domain_pts_1_1}
\end{subfigure}
\hspace*{\fill} 
\begin{subfigure}{0.45\textwidth}
    \includegraphics[height=123px, width=174px]{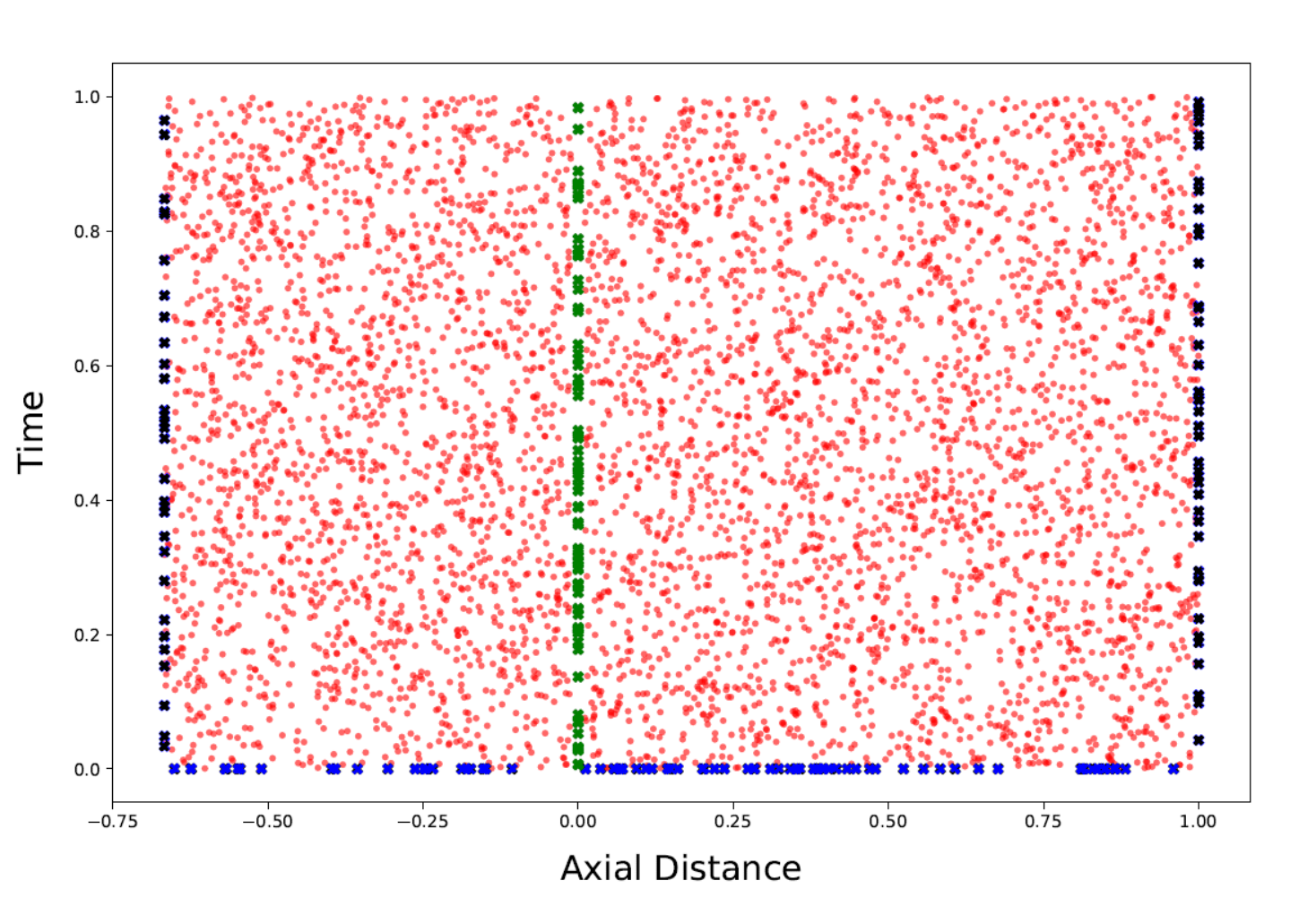}
    \caption{}
    \label{fig:domain_pts_1_2}
\end{subfigure}
\begin{subfigure}{0.45\textwidth}
        \includegraphics[height=123px, width=174px]{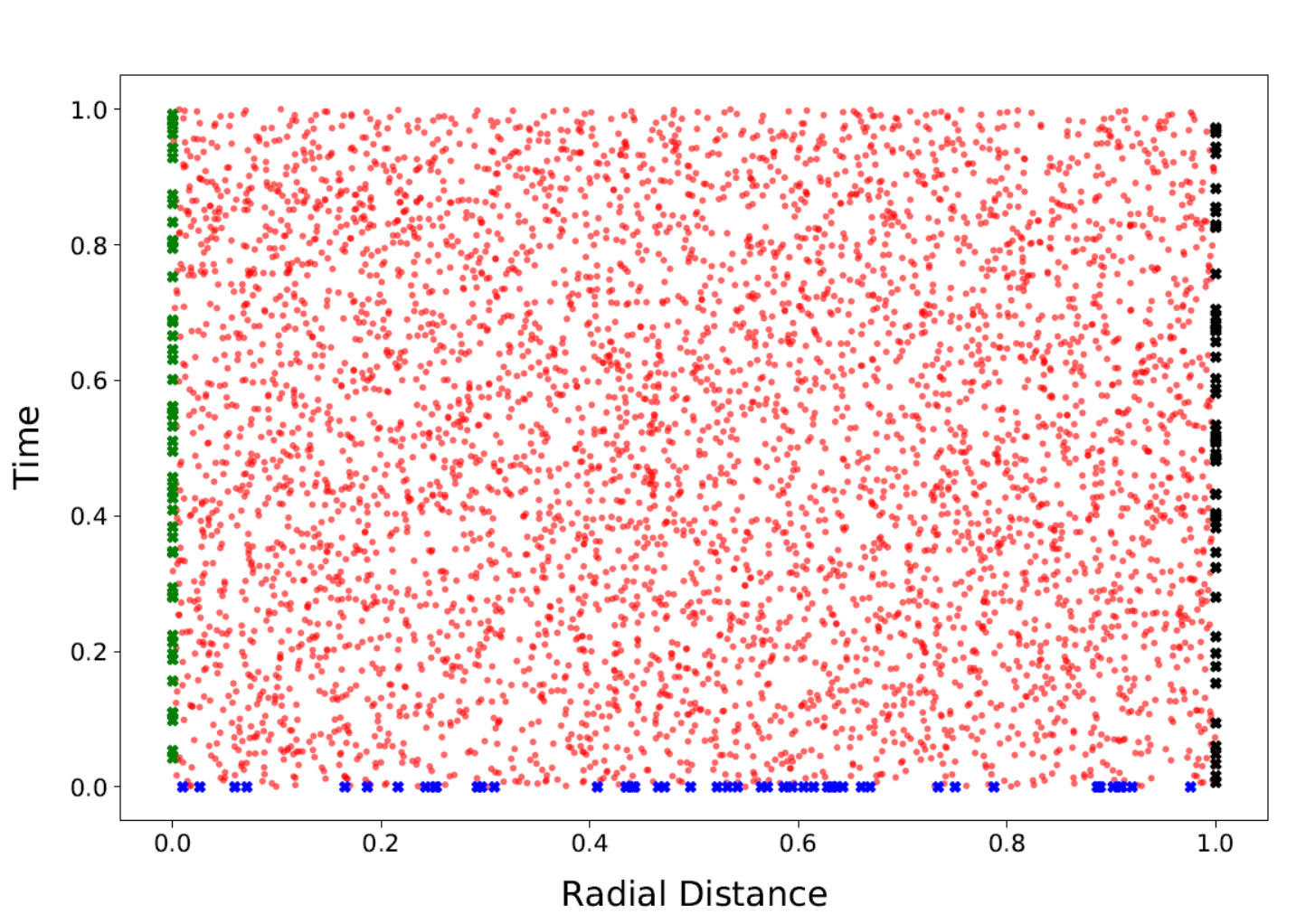}
    \caption{}
    \label{fig:domain_pts_radial}
\end{subfigure}
\caption{Domain points: (a)  Input points for 1D diffusivity PINN model with a single no-flow boundary. (b) Input points for 1D diffusivity PINN
model with two no-flow boundary conditions. (c)  Input points for the radial diffusivity PINN model with a single no-flow boundary. Red sets are for the PDE (collocation points), green sets are for
the well condition, black sets are for the
no-flow boundary conditions, and blue
sets are for the initial condition}
\label{fig:domain_pts}
\end{figure}

First, we write eq. \ref{1d_diff_eq} and eq. \ref{diff_eq_cylindrical} in a dimensionless form. Note that in this work the permeabilities $k_x$ and $k_r$ in eqs. \ref{1d_diff_eq} ,\ref{diff_eq_cylindrical} are assumed to be constant and therefore can be moved to the right hand side of the equation by division.

\begin{subequations}
\begin{align}\label{dim_1d_cart}
\begin{split}
    &\frac{\partial p_D}{\partial t_D} = \frac{\partial ^2p_D}{\partial x_D^2}\\
    & x_D = \frac{x}{x_e}; \hspace{0.3cm} t_D=\frac{k_xt}{\phi \mu c_t x_e^2}; \hspace{0.3cm} p_D=\frac{p-p_{wf}}{p_0-p_{wf}}
\end{split}\\[15pt]
\begin{split}\label{dim_radial}
    &\frac{\partial p_D}{\partial t_D} = \frac{1}{r_D}\frac{\partial p_D}{\partial r_D} + \frac{\partial ^2p_D}{\partial r_D^2}\\
    &r_D = \frac{r}{r_e}; \hspace{0.3cm} t_D=\frac{k_rt}{\phi \mu c_t r_e^2}\\
    &p_D=\frac{p-p_{wf}}{p_0-p_{wf}}\hspace{0.6cm} \text{for case 2.1}\\
    &p_D=\frac{p_0-p}{p_{ch}}\hspace{1.0cm} \text{for case 2.2}\\
\end{split}
\end{align}
\end{subequations}
\newline
where $x_e$ and $r_e$ are the maximum distances from the well for the cartesian and radial cases respectively, and the characteristic pressure $p_{ch}$ is given by:
\begin{align*}
    p_{ch}=\frac{q\mu}{2\pi kh}
\end{align*}
where $q$ is the production flow rate of the well for case 2.2. Notice that $x_D, r_D \in [0,1]$ by definition. $p_D$ for all cases apart from case 2.2 are also scaled by definition. Regarding case 2.2, in this study we work with conditions where $p_0 - p < p_{ch}$ and therefore $p_D$ for case 2.2 is also scaled by definition. Lastly, we need to scale $t_D$ and thus eq. \ref{dim_1d_cart} and eq. \ref{dim_radial} can be written as follows:
\begin{equation}\label{scaled_cart}
\begin{split}
    &\frac{1}{t_{D,max}}\frac{\partial \bar{p}}{\partial \bar{t}} = \frac{\partial ^2\bar{p}}{\partial \bar{x}^2}\\[5pt]
    &\bar{x}=x_D; \hspace{0.5cm} \bar{t}=\frac{t_D}{t_{D,max}}; \hspace{0.5cm} t_{D,max} = \frac{kt_{max}}{\phi \mu c_t x_e^2}
\end{split}
\end{equation}
\begin{equation}\label{scaled_rad}
\begin{split}
    &\frac{1}{t_{D,max}}\frac{\partial \bar{p}}{\partial \bar{t}} = \frac{1}{\bar{r}}\frac{\partial \bar{p}}{\partial \bar{r}} + \frac{\partial ^2\bar{p}}{\partial \bar{r}^2}\\[5pt]
    &\bar{r}=r_D; \hspace{0.5cm} \bar{t}=\frac{t_D}{t_{D,max}}; \hspace{0.5cm} t_{D,max} = \frac{kt_{max}}{\phi \mu c_t r_e^2}
\end{split}
\end{equation}
Now eq. \ref{scaled_cart} and eq. \ref{scaled_rad} are suitable to work with PINNs and the domain points are shown in Fig. \ref{fig:domain_pts}. Domain points are generated using random uniform sampling.

\subsection{Loss Components}
For case 1.1 the components of the loss function are as follows:
    \begin{align}
        \mathcal{L}_r(\boldsymbol{w}) =& \frac{1}{N_r} \sum_{i=1}^{N_r}r({\boldsymbol{x}^i_r}, {t^i_r}, \boldsymbol{w}),\\
        \mathcal{L}_w(\boldsymbol{w}) =& \frac{1}{N_w} \sum_{i=1}^{N_w}\big(\Tilde{u}({\boldsymbol{x}^i_w}, {t^i_w}, \boldsymbol{w})-g(\boldsymbol{x}^i_w, t^i_w)\big)^2,\\
        \mathcal{L}_b(\boldsymbol{w}) =& \frac{1}{N_b} \sum_{i=1}^{N_b}\big(\Tilde{u}({\boldsymbol{x}^i_b}, {t^i_b}, \boldsymbol{w})-g(\boldsymbol{x}^i_b, t^i_b)\big)^2,\\
        \mathcal{L}_0(\boldsymbol{w}) =& \frac{1}{N_0} \sum_{i=1}^{N_0}\big(\Tilde{u}({\boldsymbol{x}^i_0}, {0^i}, \boldsymbol{w})-u_0(\boldsymbol{x}^i_0)\big)^2,  
    \end{align}
    
where $\mathcal{L}_r$, $\mathcal{L}_w$, $\mathcal{L}_{b1}$, $\mathcal{L}_{b2}$, and $\mathcal{L}_0$ are the loss function components associated with PDE residual, well, right no-flow boundary, left no-flow boundary, and initial condition respectively.

Finally for case 2.1 and 2.2 the components of the loss function are as follows:
    \begin{align}
        \mathcal{L}_r(\boldsymbol{w}) =& \frac{1}{N_r} \sum_{i=1}^{N_r}r({\boldsymbol{x}^i_r}, {t^i_r}, \boldsymbol{w}),\\
        \mathcal{L}_w(\boldsymbol{w}) =& \frac{1}{N_w} \sum_{i=1}^{N_w}\big(\Tilde{u}({\boldsymbol{x}^i_w}, {t^i_w}, \boldsymbol{w})-g(\boldsymbol{x}^i_w, t^i_w)\big)^2,\\
        \mathcal{L}_{b1}(\boldsymbol{w}) =& \frac{1}{N_{b1}} \sum_{i=1}^{N_{b1}}\big(\Tilde{u}({\boldsymbol{x}^i_{b1}}, {t^i_{b1}}, \boldsymbol{w})-g(\boldsymbol{x}^i_{b1}, t^i_{b1})\big)^2,\\
        \mathcal{L}_{b2}(\boldsymbol{w}) =& \frac{1}{N_{b2}} \sum_{i=1}^{N_{b2}}\big(\Tilde{u}({\boldsymbol{x}^i_{b2}}, {t^i_{b2}}, \boldsymbol{w})-g(\boldsymbol{x}^i_{b2}, t^i_{b2})\big)^2,\\
        \mathcal{L}_0(\boldsymbol{w}) =& \frac{1}{N_0} \sum_{i=1}^{N_0}\big(\Tilde{u}({\boldsymbol{x}^i_0}, {0^i}, \boldsymbol{w})-u_0(\boldsymbol{x}^i_0)\big)^2,  
    \end{align}

where $\mathcal{L}_r$, $\mathcal{L}_w$, $\mathcal{L}_{b1}$, $\mathcal{L}_{b2}$, and $\mathcal{L}_0$ are the loss function components associated with PDE residual, well, right no-flow boundary, left no-flow boundary, and initial condition respectively.

Finally for case 2.1 and 2.2 the components of the loss function are as follows:
    \begin{align}
        \mathcal{L}_r(\boldsymbol{w}) =& \frac{1}{N_r} \sum_{i=1}^{N_r}r({\boldsymbol{r}^i_r}, {t^i_r}, \boldsymbol{w}),\\
        \mathcal{L}_w(\boldsymbol{w}) =& \frac{1}{N_w} \sum_{i=1}^{N_w}\big(\Tilde{u}({\boldsymbol{r}^i_w}, {t^i_w}, \boldsymbol{w})-g(\boldsymbol{r}^i_w, t^i_w)\big)^2,\\
        \mathcal{L}_b(\boldsymbol{w}) =& \frac{1}{N_b} \sum_{i=1}^{N_b}\big(\Tilde{u}({\boldsymbol{r}^i_b}, {t^i_b}, \boldsymbol{w})-g(\boldsymbol{r}^i_b, t^i_b)\big)^2,\\
        \mathcal{L}_0(\boldsymbol{w}) =& \frac{1}{N_0} \sum_{i=1}^{N_0}\big(\Tilde{u}({\boldsymbol{r}^i_0}, {0^i}, \boldsymbol{w})-u_0(\boldsymbol{r}^i_0)\big)^2,  
    \end{align}
where $\mathcal{L}_r$, $\mathcal{L}_w$, $\mathcal{L}_b$, and $\mathcal{L}_0$ are the loss function components associated with PDE residual, well, no-flow boundary, and initial condition respectively.

\section{Results}
\subsection{1D Forward Problems:}
\begin{itemize}
    \item \textbf{Case 1.1:}\\
    In case 1.1, a well producing at constant BHP and a no-flow boundary. The results are shown in Fig. \ref{fig:pres_pred_1_1a}. Once the PINN is trained production rates can be easily calculated using Darcy's law:
    \begin{align*}
        \tilde{q}=\frac{kA}{\mu}\frac{d\tilde{p}}{dx}
    \end{align*}\\
    where $A$ is flow surface area, and $\frac{d\tilde{p}}{dx}$ is the pressure gradient which can be obtained using back-propagation. Note that this is a one-dimensional problem, therefore, the flow area $A$ is set to unity. In Fig. \ref{fig:pres_pred_1_1} notice that PINNs predictions for both pressure and flow rates are continuous in time whereas the high-fidelity are discrete and user-specified. In other words, the user must specify at which specific times they wish to obtain solutions for pressures and flow rates.
    \begin{figure}[H]
    \centering
    \begin{subfigure}{0.45\textwidth}
        \includegraphics[height=128px, width=186px]{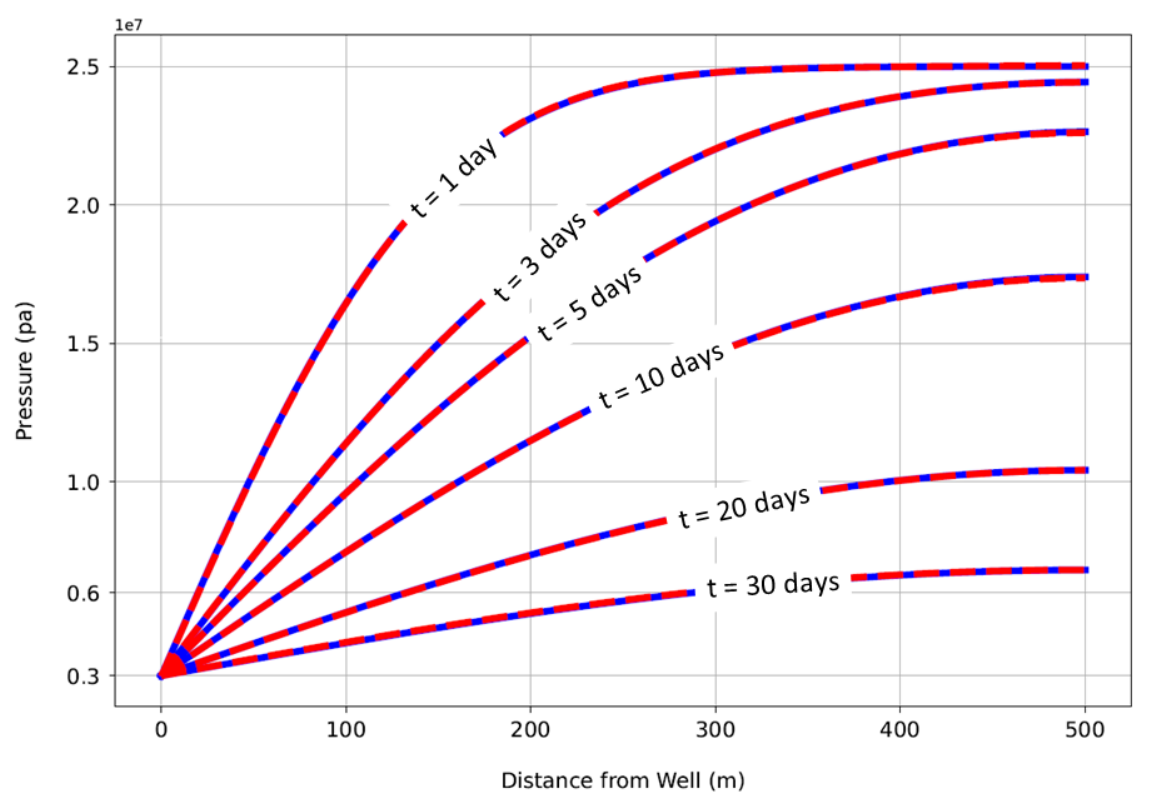}
        \caption{}
        \label{fig:pres_pred_1_1a}
    \end{subfigure}
    \hspace*{\fill}
    \begin{subfigure}{0.45\textwidth}
    \includegraphics[height=128px, width=186px]
        {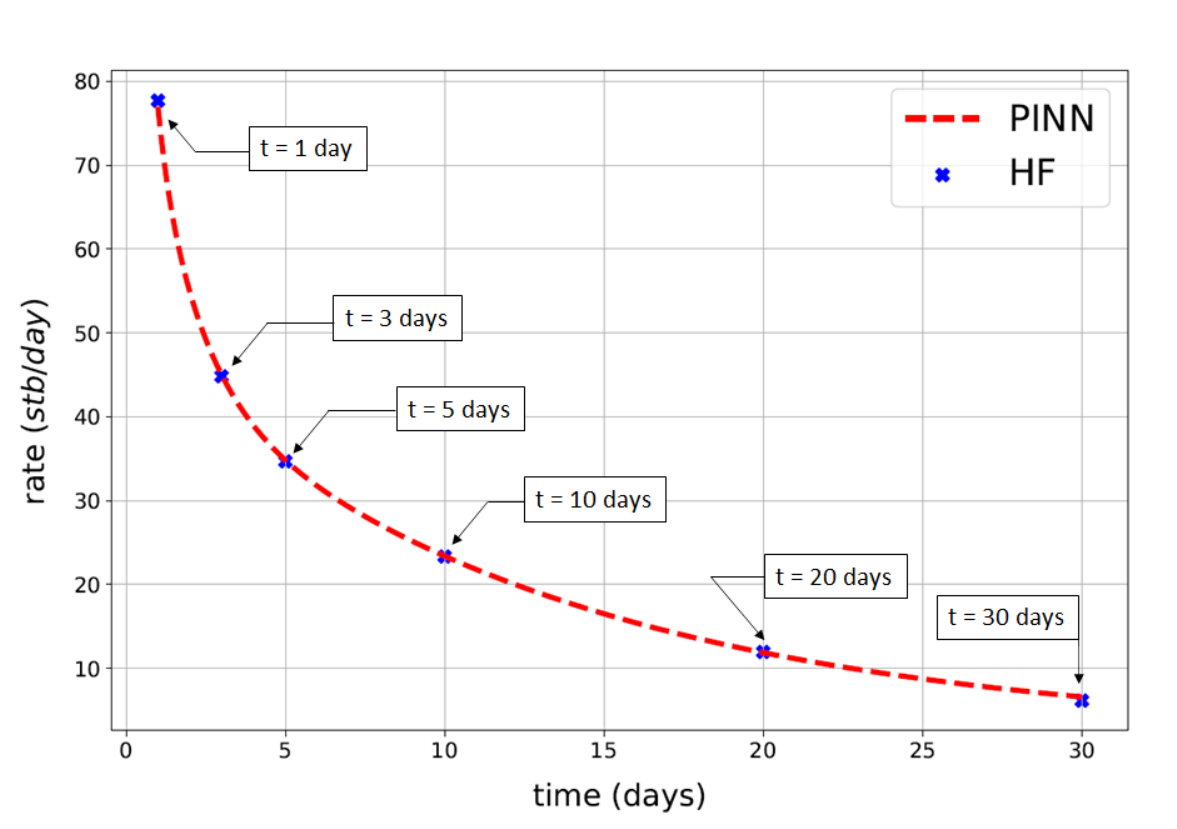}
        \caption{}
    \label{fig:pres_pred_1_1b}
    \end{subfigure}
    \caption{(a) - PINN prediction for pressure (red dashed) versus high-fidelity numerical solution (solid blue). (b) - flow rates prediction using PINN (red dashed) versus high-fidelity numerical solution (blue dots). The relative L2 error is less than 0.1\%.}
    \label{fig:pres_pred_1_1}
    \end{figure}

\item \textbf{Case 1.2:}\\
    In case 1.2 a well producing at constant BHP between two no-flow boundaries. The results are shown in Fig. \ref{fig:1d_fig_1.2}.
    \begin{figure}[H]
    \centering
    \includegraphics[height=144px, width=231px]{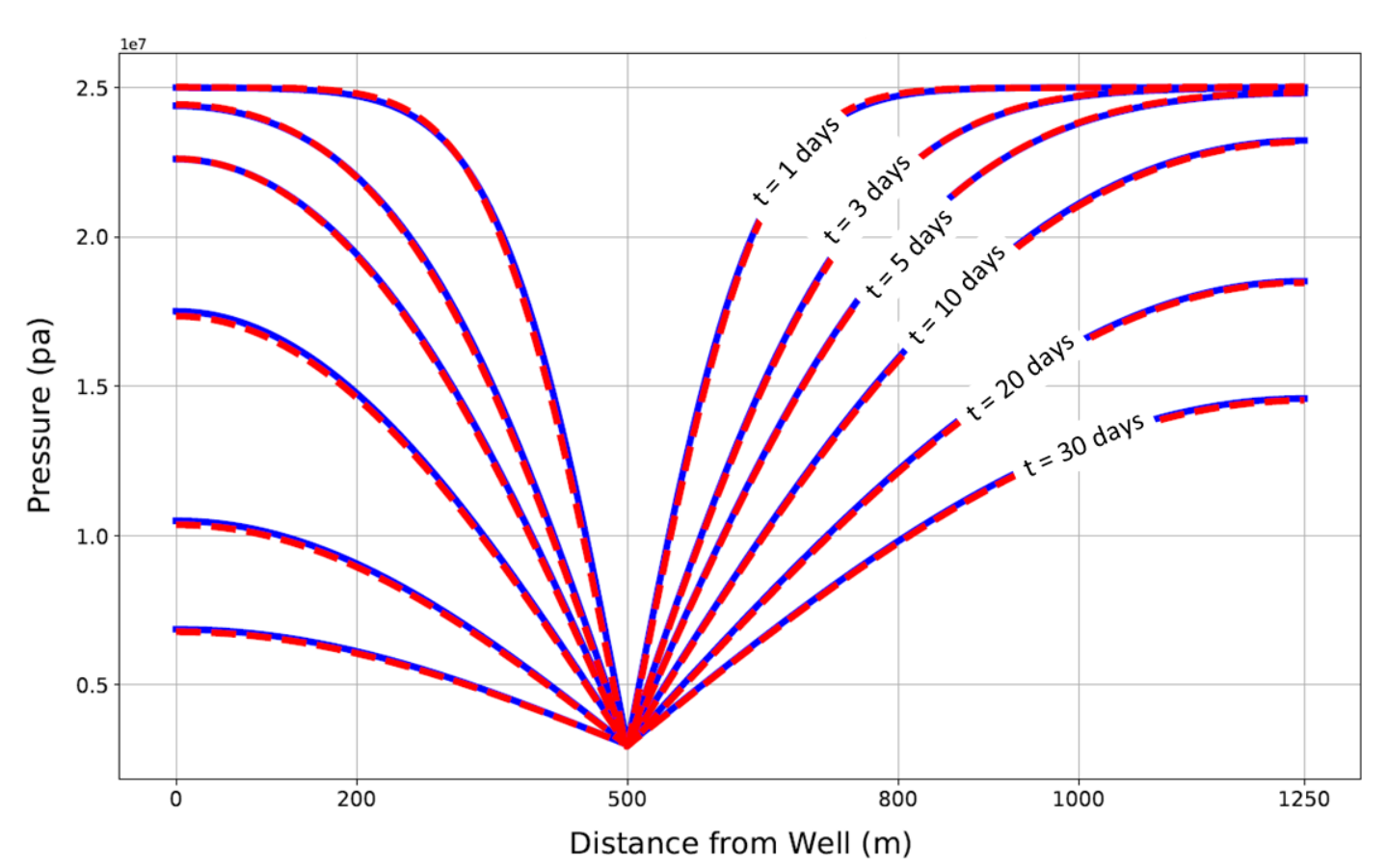}
    \caption{PINN pressure prediction (red dashed) versus high-fidelity numerical solution (solid blue). The relative L2 error is less than 0.1\%}
    \label{fig:1d_fig_1.2}
    \end{figure}
\end{itemize}
\subsection{Dimensionless Variables}
It is important to note that working with scaled dimensionless form of the PDE enables making instantaneous predictions for all reservoir properties. Any change in reservoir properties $(\phi, \mu, k, c_t, x_e)$ is equivalent to change in the dimensionless time $t_D$. Unlike numerical methods where any change in the reservoir properties requires restarting the method from the beginning, in PINNs there is no need to retrain the model since predicting for a different reservoir property is equivalent to predicting for a different dimensionless time.

\begin{figure}[H]
    \centering
    \includegraphics[height=156px, width=373px]{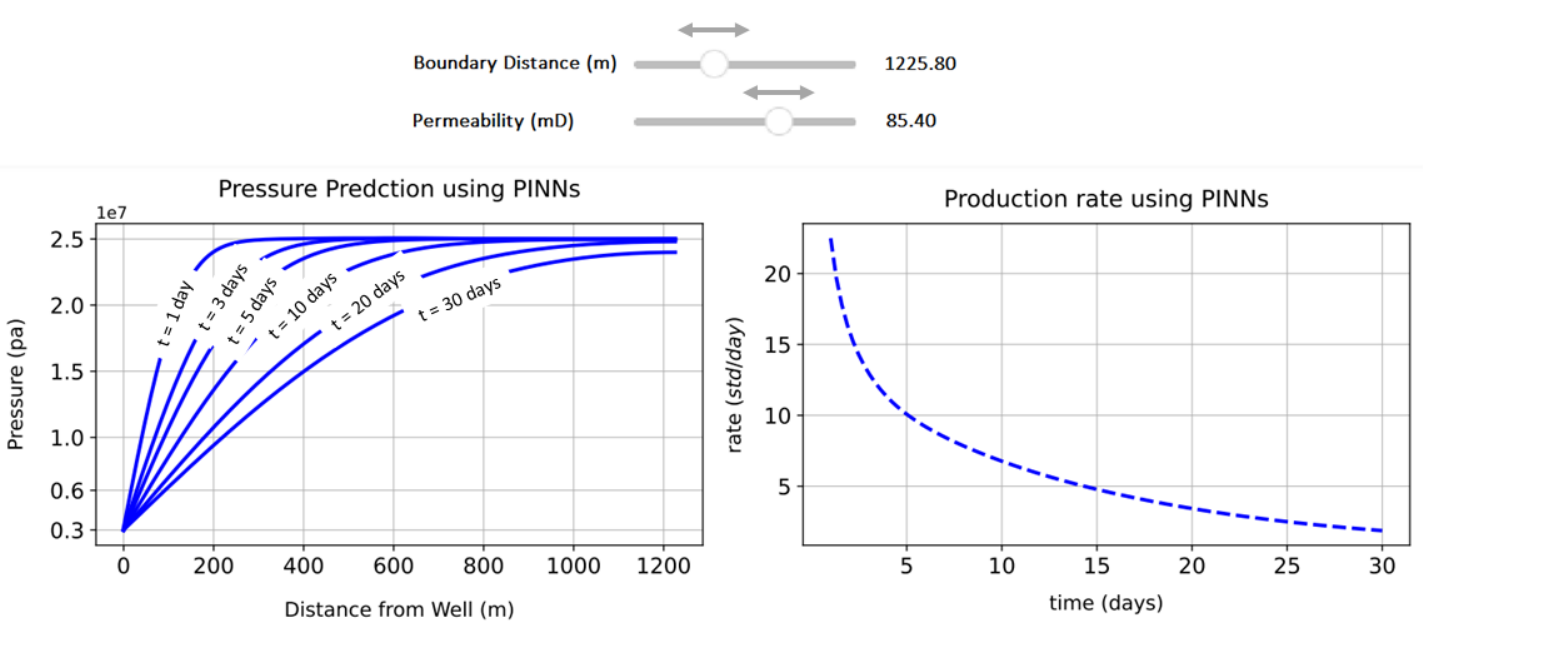}
    \caption{Interactive plot for permeability $k$ and boundary distance from well $x_e$. As we change the values with the sliders the plots are instantaneously updated with the new values.}
    \label{fig:1d_interactive}
    \end{figure}

For practical purposes we can generate interactive-plots with sliders for instantaneous predictions as in Fig. \ref{fig:1d_interactive}. Additionally, if we have production rates data placed on the right plot of Fig. \ref{fig:1d_interactive}, we can change reservoir properties until achieving a good match and with that we can estimate reservoir properties associated with the production rates data.
    
\subsection{1D Inverse Problems:}
\begin{figure}[H]
    \centering
    \includegraphics[height=148px, width=221px]{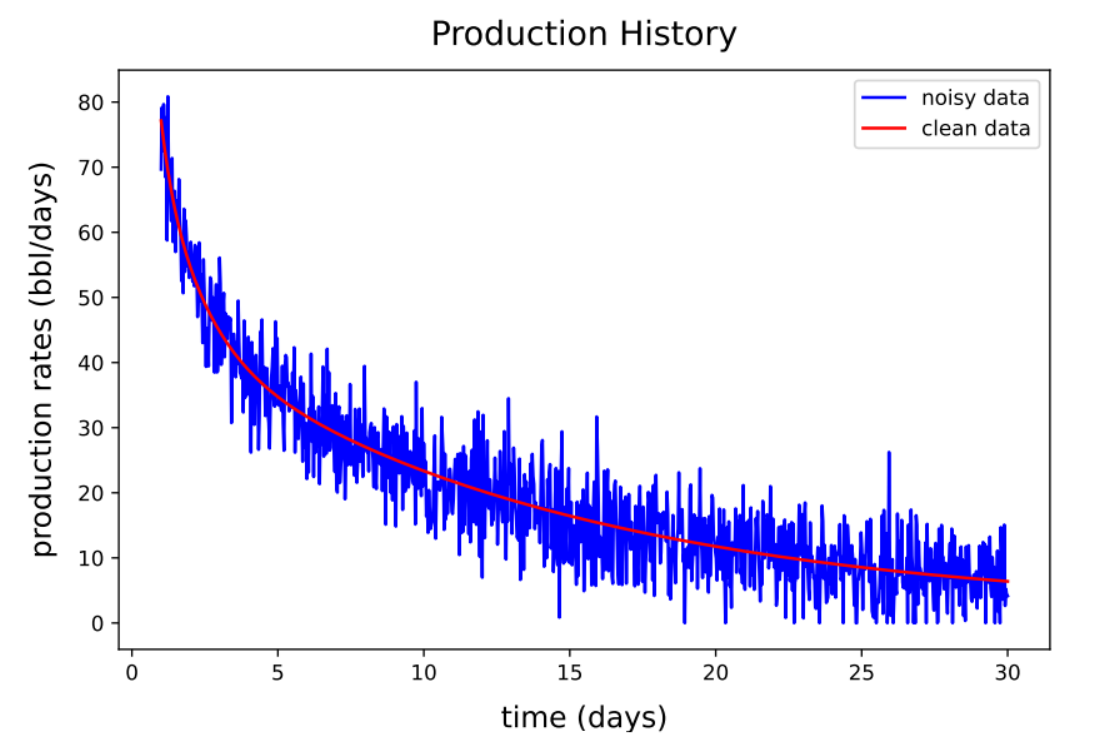}
    \caption{Production flow rates history}
    \label{fig:flow_rates}
    \end{figure}

    In this section we test the inverse capabilities of PINNs. The problem setup is the following. We introduce a production flow rate history data $q_{data}$ and we treat one of the reservoir properties as an unknown. We tackle two problems in this section, the first, we treat the permeability $k_x$ as an unknown and the second we treat the distance from the boundary $x_e$ as unknown.
    For the first problem, eq. \ref{dim_1d_cart} now has two unknowns, the pressure and the permeability with one additional boundary condition is introduced that is the flow rates history data shown in Fig. \ref{fig:flow_rates}. The loss function components now are as follows:
    
    \begin{align}\label{eq:4.1a}
        \mathcal{L}_r(\boldsymbol{w}) =& \frac{1}{N_r} \sum_{i=1}^{N_r}r({\boldsymbol{x}^i_r}, {t^i_r}, \boldsymbol{w}),\\
        \mathcal{L}_w(\boldsymbol{w}) =& \frac{1}{N_w} \sum_{i=1}^{N_w}\big(\Tilde{u}({\boldsymbol{x}^i_w}, {t^i_w}, \boldsymbol{w})-g(\boldsymbol{x}^i_w, t^i_w)\big)^2,\\
        \mathcal{L}_b(\boldsymbol{w}) =& \frac{1}{N_b} \sum_{i=1}^{N_b}\big(\Tilde{u}({\boldsymbol{x}^i_b}, {t^i_b}, \boldsymbol{w})-g(\boldsymbol{x}^i_b, t^i_b)\big)^2,\\
        \mathcal{L}_0(\boldsymbol{w}) =& \frac{1}{N_0} \sum_{i=1}^{N_0}\big(\Tilde{u}({\boldsymbol{x}^i_0}, {0^i}, \boldsymbol{w})-u_0(\boldsymbol{x}^i_0)\big)^2,\\
        \mathcal{L}_d(\boldsymbol{w}) =& \frac{1}{N_d} \sum_{i=1}^{N_d}\big(\Tilde{q}({{0}^i}, {t^i_d}, \boldsymbol{w})-q_d(t^i_d)\big)^2,\label{4.1e}
    \end{align}
    
\begin{figure}[H]
    \centering
    \includegraphics[height=148px, width=221px]{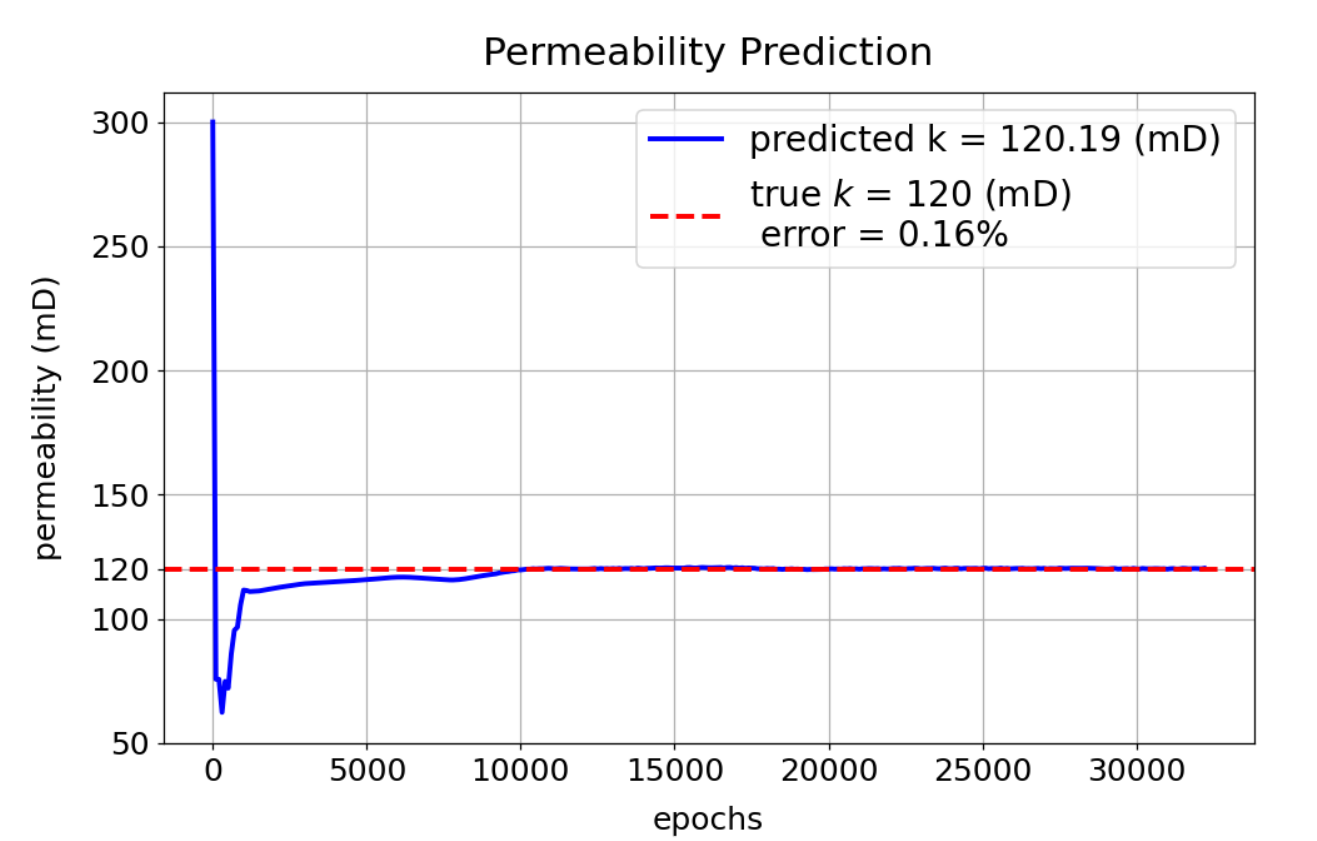}
    \caption{Permeability prediction for a given production flow rates history data.}
    \label{fig:perm_pred}
    \end{figure}

\begin{figure}[H]
    \centering
    \includegraphics[height=148px, width=221px]{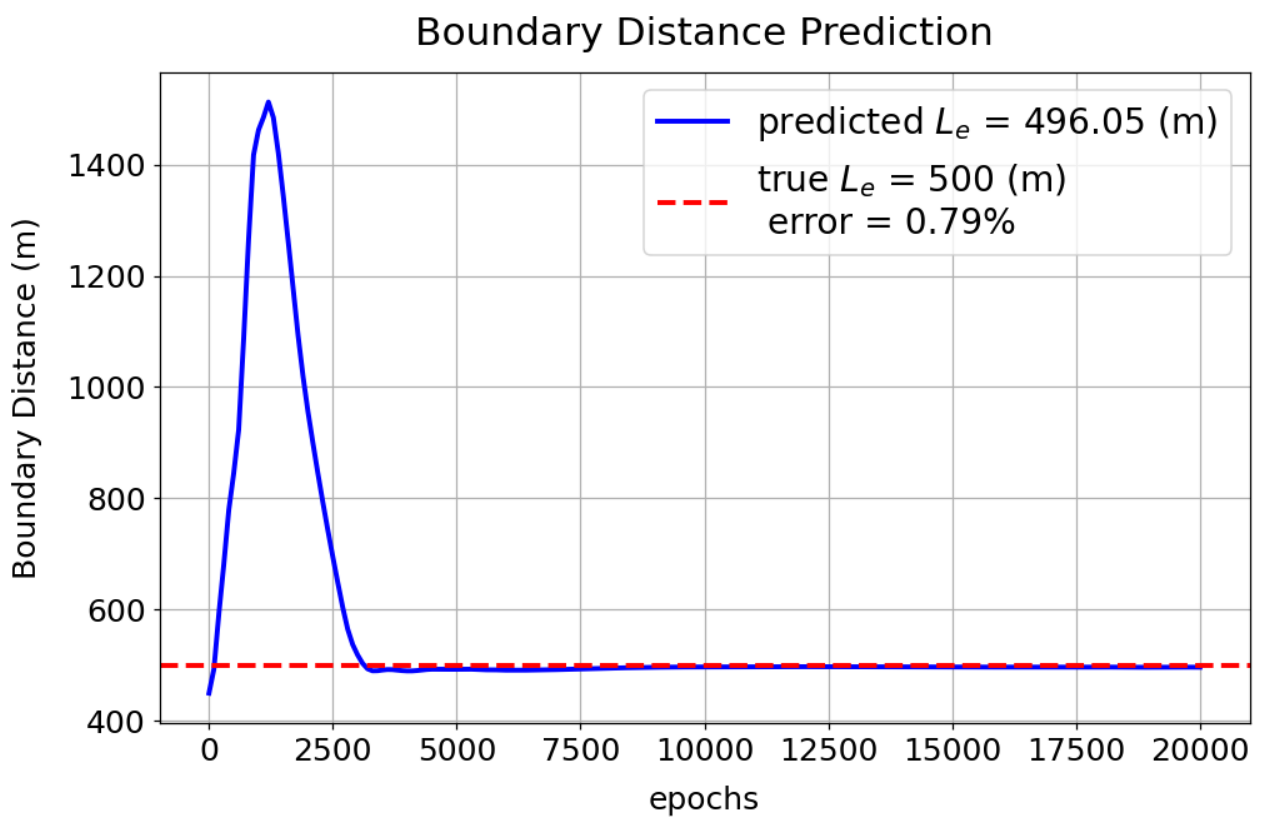}
    \caption{Boundary distance prediction for a given production flow rates history data.}
    \label{fig:boundary_pred}
    \end{figure}

    Where $\Tilde{q}$ in eq. \ref{4.1e} is the predicted flow rates by the PINN and $q_d$ is the production flow rates data. As shown in Fig. \ref{fig:perm_pred}, a PINN model was able to learn the unknown permeability that satisfies the PDE, the initial condition, boundary conditions and the production history data with a relative error of 0.16\%. Similarly, for the second problem with unknown boundary distance, a PINN model is trained to satisfy the PDE, the initial condition, boundary conditions and the production history data with a relative error of 0.79\% as shown in Fig. \ref{fig:boundary_pred}. It is important to note that both models were trained with the noisy data (blue curve) in Fig. \ref{fig:flow_rates}.\\
    In order for the model to learn an unknown reservoir property, the property is learnt by gradient descent similar to network weights and biases. For the case with unknown permeability we rewrite eq. \ref{1d_diff_eq}:
    
    \begin{align}
    \begin{split}
        &\frac{\phi \mu c_t}{k_x} \bigg(\frac{k_m}{k_m}\bigg) \frac{\partial p}{\partial t} = \frac{\partial^2 p}{\partial x^2}\\\\
        &x \in [0, x_e], \hspace{0.3cm} t\in[0, t_f]
    \end{split}
    \end{align}
    
    where $k_m$ is an upper estimation of the permeability field and $k_x$ is the unknown permeability. Here we take $k_m=300[mD]$.
    
    \begin{align}
    \begin{split}
        &\frac{\phi \mu c_t}{k_m} \bigg(\frac{k_m}{k_x}\bigg) \frac{\partial p}{\partial t} = \frac{\partial^2 p}{\partial x^2}\\\\
        &\eta=\frac{k_x}{k_m}\\\\
        &\frac{\phi \mu c_t}{k_m} \frac{\partial p}{\partial t} = \eta\frac{\partial^2 p}{\partial x^2}
    \end{split}
    \end{align}
    
     where $\eta$ is called the permeability factor. Now we can rewrite eq. \ref{dim_1d_cart} as follows:
    
    \begin{align}\label{dim_1d_cart_interactive}
    \begin{split}
    & \frac{1}{t_{D,max}} \frac{\partial p_D}{\partial t_D} =\eta \frac{\partial ^2p_D}{\partial x_D^2}\\
    & x_D = \frac{x}{x_e}; \hspace{0.3cm} t_D=\frac{k_mt}{\phi \mu c_t x_e^2}; \hspace{0.3cm} p_D=\frac{p-p_{wf}}{p_0-p_{wf}}
    \end{split}
    \end{align}
    
    In addition to the previous, notice now that the residual of the PDE of eq. \ref{dim_1d_cart_interactive} is also a function of $\eta$, and the loss function of the residual can be written as follows:
    
    \begin{align}\label{residual_loss_eta}
        \mathcal{L}_r(\boldsymbol{w}, \eta) =& \frac{1}{N_r} \sum_{i=1}^{N_r}r({\boldsymbol{x}^i_r}, {t^i_r}, \boldsymbol{w}, \eta)
    \end{align}
    
    The production flow rates history data $q_{data}$ is also a function of $\eta$ for the following reason:
    \begin{align*}
        &q_d=\frac{kA}{\mu}\frac{dp}{dx} = \frac{kA}{\mu}\frac{k_m}{k_m}\frac{dp}{dx}=
        \frac{k_mA}{\mu}\eta\frac{dp}{dx}=
        \eta\frac{k_mA}{\mu}\bigg(\frac{p_0-p_{wf}}{x_e}\bigg)\frac{dp_D}{dx_D}
    \end{align*}
    
    \begin{align}
       \underbrace{\frac{q_d\mu}{k_mA}\bigg(\frac{x_e}{p_0-p_{wf}}\bigg)}_\text{known quantity}\hspace{0.5cm}=\hspace{0.5cm}\underbrace{\eta}_\text{learnt quantity} \hspace{0.5cm} \times \hspace{0.5cm}\underbrace{\frac{dp_D}{dx_D}}_\text{back propagation}
    \end{align}
    
    \begin{align}
       \mathcal{L}_d(\boldsymbol{w}, \eta) = \frac{q_d\mu}{k_mA}\bigg(\frac{x_e}{p_0-p_{wf}}\bigg)\hspace{0.1cm}-\hspace{0.1cm} \eta \hspace{0.1cm} \times \hspace{0.1cm}\frac{dp_D}{dx_D}
    \end{align}
    The total loss function of the network is therefore a function of $\eta$ and given by:
    \begin{align}\label{total_loss_eta}
        \mathcal{L}(\boldsymbol{w}, \eta) = \mathcal{L}_r(\boldsymbol{w}, \eta)+ \mathcal{L}_0(\boldsymbol{w})+ \mathcal{L}_w(\boldsymbol{w})+ \mathcal{L}_b(\boldsymbol{w})+
        \mathcal{L}_d(\boldsymbol{w}, \eta)
    \end{align}
    
    Similar analogy can be applied for the case with unknown boundary distance. The results for both cases are shown in Fig. \ref{fig:perm_pred} and Fig. \ref{fig:boundary_pred}.
    
To summarize this section, we showed that from production data we are able to estimate the missing permeability and distance from sealing boundary without having to shut-in the well, whereas, in buildup tests to estimate these properties a well shut-in is required.
\subsection{Forward 2D case}
\begin{figure}[!ht]
    \centering
    \includegraphics[height=148px, width=150px]{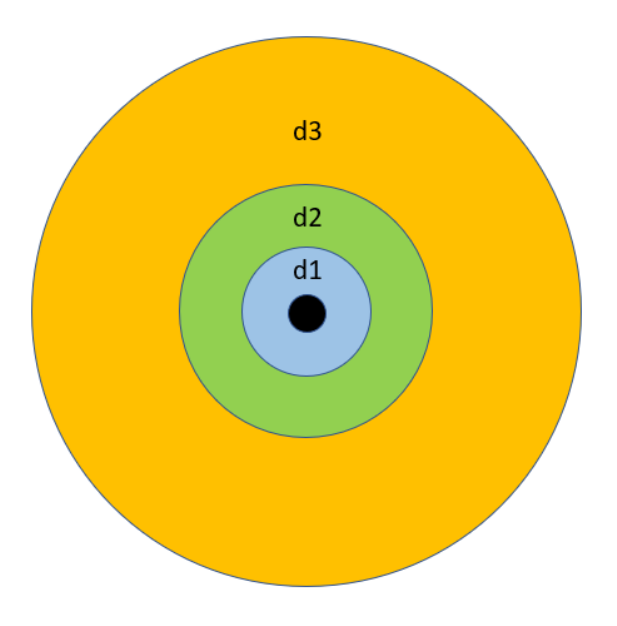}
    \caption{Decomposition of circular reservoir into 3 subdomains. The reservoir has a a well (black) in the center and no-flow outer boundary .}
    \label{fig:subdomains}
\end{figure}

To this end, we dealt with the 1D diffusivity equation which is not a stiff PDE, and we saw how a single physics-informed neural network was able to learn the solution. When working with the 2D diffusivity equation which is a very stiff PDE (especially close to the well) the problem become much more challenging. For several traditional neural network architectures that we tried, a single neural network was not able to learn the solution for the 2D case. A possible reason for this is because the solution of the equation exhibits highly logarithmic nature close to the well, and linear nature very far from the well with a transition nature in between. Applying the conservative PINNs (cPINNs) methodology \citep{cPINNs} to a circular reservoir with well in the center, we decompose the reservoir into multiple subdomains and each subdomain is associated with a completely independent neural network, and at the interface between the subdomains we require the pressures predictions and pressure gradients predictions (fluxes) to be equal.\\
As an example let's take Fig. \ref{fig:subdomains}. For domain No.1, the loss function components are:

    \begin{align*}
        \mathcal{L}_r^{(1)}\big(\boldsymbol{w_1}\big) =& \frac{1}{N_r} \sum_{i=1}^{N_r}r^{(1)}({\boldsymbol{x}^i_r}, {t^i_r}, \boldsymbol{w_1})^2,\\
        \mathcal{L}_0^{(1)}\big(\boldsymbol{w_1}\big) =& \frac{1}{N_0} \sum_{i=1}^{N_0}\bigg(\Tilde{u}^{(1)}({\boldsymbol{x}^i_0}, {0^i}, \boldsymbol{w_1})-u_0(\boldsymbol{x}^i_0)\bigg)^2,\\
        \mathcal{L}_w^{(1)}\big(\boldsymbol{w_1}\big) =& \frac{1}{N_w} \sum_{i=1}^{N_w}\bigg(\Tilde{u}^{(1)}({\boldsymbol{x}^i_w}, {t^i_w}, \boldsymbol{w_1})-g(\boldsymbol{x}^i_w, t^i_w)\bigg)^2,\\
        \mathcal{L}_{pc,12}^{(1)}\big(\boldsymbol{w_1}, \boldsymbol{w_2}\big) =& \frac{1}{N_c} \sum_{i=1}^{N_c}\bigg(\Tilde{u}^{(1)}\big({\boldsymbol{x}^i_{c,1}}, {t^i_{c,1}}, \boldsymbol{w_1}\big)-\Tilde{u}^{(2)}\big({\boldsymbol{x}^i_{c,2}}, {t^i_{c,2}}, \boldsymbol{w_2}\big)\bigg)^2,\\
        \mathcal{L}_{fc,12}^{(1)}\big(\boldsymbol{w_1}, \boldsymbol{w_2}\big) =& \frac{1}{N_c} \sum_{i=1}^{N_c}\bigg(\Tilde{u}_x^{(1)}\big({\boldsymbol{x}^i_{c,1}}, {t^i_{c,1}}, \boldsymbol{w_1}\big)-\Tilde{u}_x^{(2)}\big({\boldsymbol{x}^i_{c,2}}, {t^i_{c,2}}, \boldsymbol{w_2}\big)\bigg)^2,\\
        \mathcal{L}^{(1)}\big(\boldsymbol{w_1}, \boldsymbol{w_2}\big) =&\mathcal{L}_r^{(1)} + \mathcal{L}_0^{(1)} + \mathcal{L}_w^{(1)} + \mathcal{L}_{pc,12}^{(1)} + \mathcal{L}_{fc,12}^{(1)}
    \end{align*}

Domain No.2, the loss function components are:

\begin{align*}
    \mathcal{L}_r^{(2)}\big(\boldsymbol{w_2}\big) =& \frac{1}{N_r} \sum_{i=1}^{N_r}r^{(2)}({\boldsymbol{x}^i_r}, {t^i_r}, \boldsymbol{w_2})^2,\\
    \mathcal{L}_0^{(2)}\big(\boldsymbol{w_2}\big) =& \frac{1}{N_0} \sum_{i=1}^{N_0}\bigg(\Tilde{u}^{(2)}({\boldsymbol{x}^i_0}, {0^i}, \boldsymbol{w_2})-u_0(\boldsymbol{x}^i_0)\bigg)^2,\\
    \mathcal{L}_{pc,21}^{(2)}\big(\boldsymbol{w_1}, \boldsymbol{w_2}\big) =& \frac{1}{N_c} \sum_{i=1}^{N_c}\bigg(\Tilde{u}^{(1)}\big({\boldsymbol{x}^i_{c,1}}, {t^i_{c,1}}, \boldsymbol{w_1}\big)-\Tilde{u}^{(2)}\big({\boldsymbol{x}^i_{c,2}}, {t^i_{c,2}}, \boldsymbol{w_2}\big)\bigg)^2,\\
    \mathcal{L}_{fc,21}^{(2)}\big(\boldsymbol{w_1}, \boldsymbol{w_2}\big) =& \frac{1}{N_c} \sum_{i=1}^{N_c}\bigg(\Tilde{u}_x^{(1)}\big({\boldsymbol{x}^i_{c,1}}, {t^i_{c,1}}, \boldsymbol{w_1}\big)-\Tilde{u}_x^{(2)}\big({\boldsymbol{x}^i_{c,2}}, {t^i_{c,2}}, \boldsymbol{w_2}\big)\bigg)^2,\\
    \mathcal{L}_{pc,23}^{(2)}\big(\boldsymbol{w_2}, \boldsymbol{w_3}\big) =& \frac{1}{N_c} \sum_{i=1}^{N_c}\bigg(\Tilde{u}^{(2)}\big({\boldsymbol{x}^i_{c,2}}, {t^i_{c,2}}, \boldsymbol{w_2}\big)-\Tilde{u}^{(3)}\big({\boldsymbol{x}^i_{c,3}}, {t^i_{c,3}}, \boldsymbol{w_3}\big)\bigg)^2,\\
    \mathcal{L}_{fc,23}^{(2)}\big(\boldsymbol{w_2}, \boldsymbol{w_3}\big) =& \frac{1}{N_c} \sum_{i=1}^{N_c}\bigg(\Tilde{u}_x^{(2)}\big({\boldsymbol{x}^i_{c,2}}, {t^i_{c,2}}, \boldsymbol{w_2}\big)-\Tilde{u}_x^{(3)}\big({\boldsymbol{x}^i_{c,3}}, {t^i_{c,3}}, \boldsymbol{w_3}\big)\bigg)^2,\\[10pt]
    \mathcal{L}^{(1)}\big(\boldsymbol{w_1}, \boldsymbol{w_2}, \boldsymbol{w_3}\big) =&\mathcal{L}_r^{(2)} + \mathcal{L}_0^{(2)} + \mathcal{L}_{pc,21}^{(2)} + \mathcal{L}_{fc,21}^{(2)} + \mathcal{L}_{pc,23}^{(2)} + \mathcal{L}_{fc,23}^{(2)}
\end{align*}

\begin{figure}[!ht]
    \centering
    \includegraphics[height=114px, width=385px]{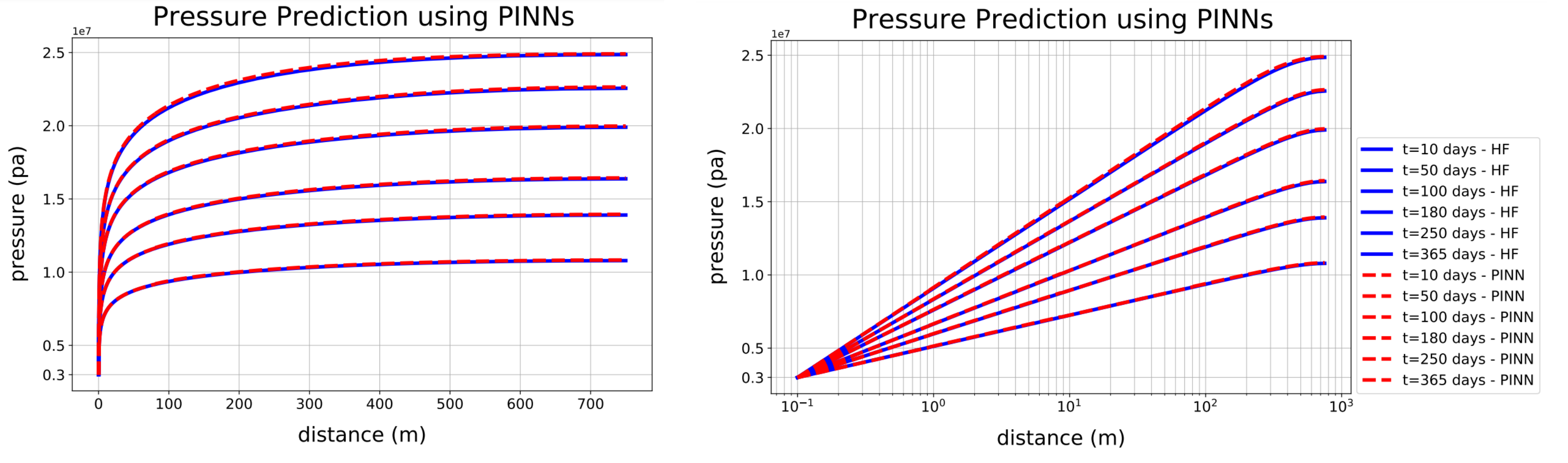}
    \caption{Pressure predictions using the decomposition method for a circular reservoir with a well in the center controlled by constant BHP and a no-flow outer boundary versus high-fidelity (HF) numerical solution. The left figure is linear-linear and the right figure is semi-log. The relative L2 error is less than 0.2\%}
    \label{fig:const_pres_2D}
\end{figure}

\begin{figure}[!ht]
    \centering
    \includegraphics[height=114px, width=385px]{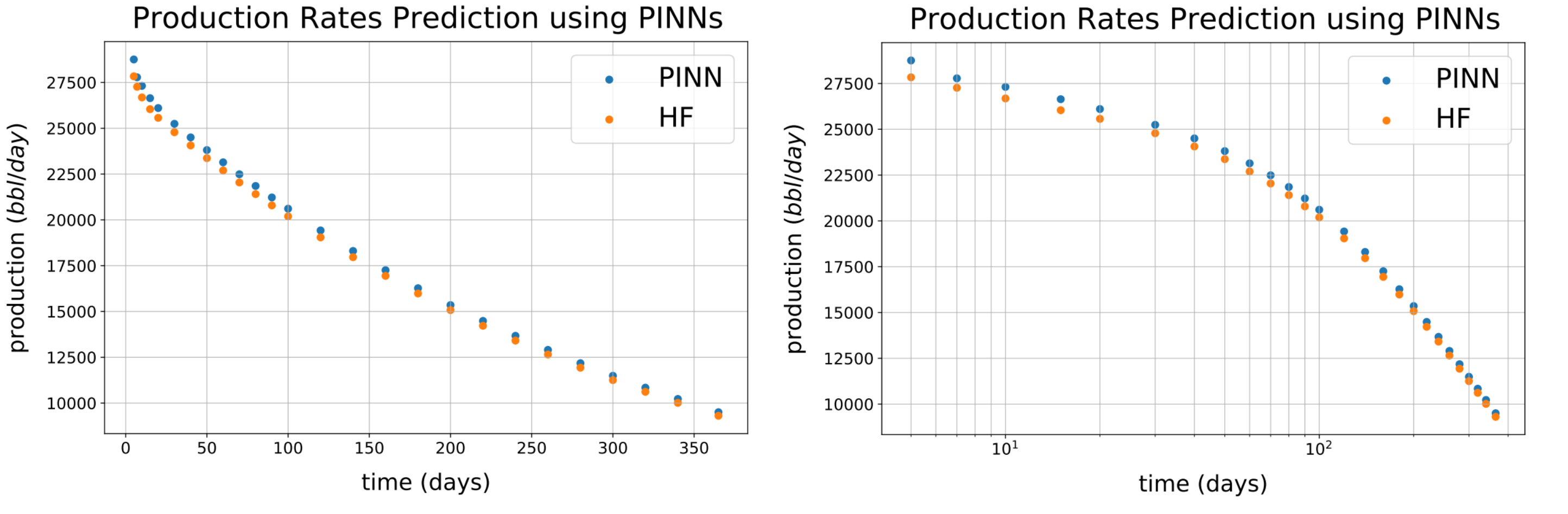}
    \caption{Production rates predictions using the decomposition method for a circular reservoir with a well in the center controlled by constant BHP and a no-flow outer boundary versus high-fidelity (HF) numerical solution. The left figure is linear-linear and the right figure is semi-log. The relative L2 error is less than 1.5\%}
    \label{fig:rates_2D}
\end{figure}

\begin{figure}[H]
    \centering
    \includegraphics[height=114px, width=385px]{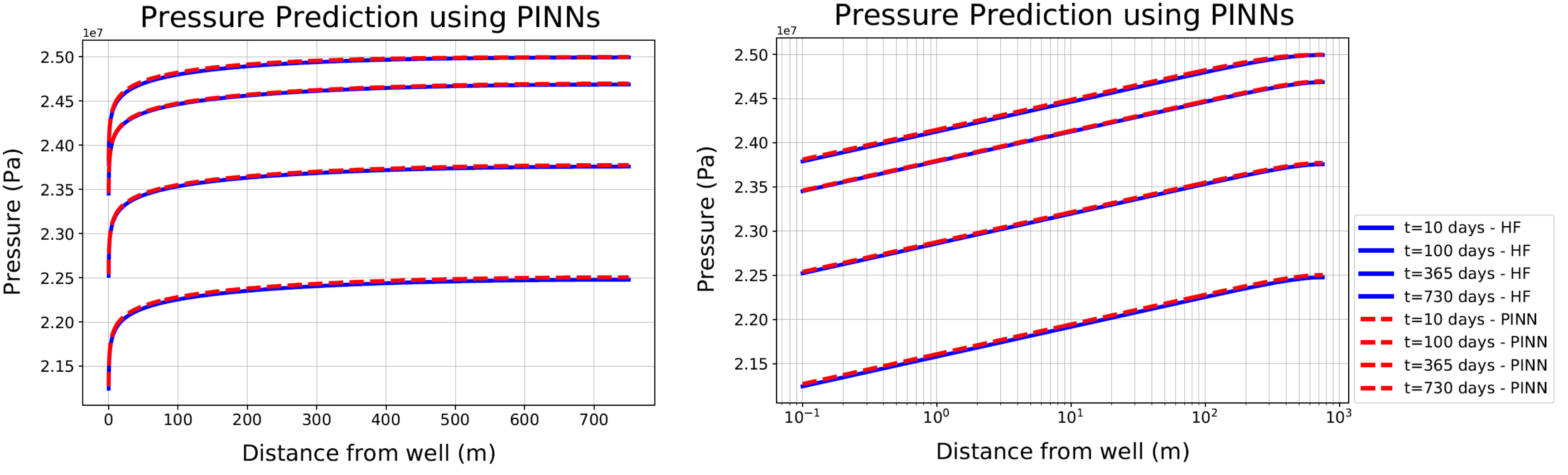}
    \caption{Pressure predictions using the decomposition method for a circular reservoir with a well in the center controlled by constant production rate and a no-flow outer boundary versus high-fidelity (HF) numerical solution. The left figure is linear-linear and the right figure is semi-log. The relative L2 error is less than 0.2\%}
    \label{fig:const_pres_rates_pred_2D}
\end{figure}

Domain No.3, the loss function components are:

\begin{align*}
    \mathcal{L}_r^{(3)}\big(\boldsymbol{w_3}\big) =& \frac{1}{N_r} \sum_{i=1}^{N_r}r^{(3)}({\boldsymbol{x}^i_r}, {t^i_r}, \boldsymbol{w_3})^2,\\
    \mathcal{L}_0^{(3)}\big(\boldsymbol{w_3}\big) =& \frac{1}{N_0} \sum_{i=1}^{N_0}\bigg(\Tilde{u}^{(3)}({\boldsymbol{x}^i_0}, {0^i}, \boldsymbol{w_3})-u_0(\boldsymbol{x}^i_0)\bigg)^2,\\
    \label{eq:4.3.3c}
    \mathcal{L}_b^{(3)}\big(\boldsymbol{w_3}\big) =& \frac{1}{N_b} \sum_{i=1}^{N_b}\bigg(\Tilde{u}^{(3)}({\boldsymbol{x}^i_b}, {t^i_b}, \boldsymbol{w_3})-g(\boldsymbol{x}^i_b, t^i_b)\bigg)^2,\\
    \mathcal{L}_{pc,32}^{(1)}\big(\boldsymbol{w_3}, \boldsymbol{w_2}\big) =& \frac{1}{N_c} \sum_{i=1}^{N_c}\bigg(\Tilde{u}^{(3)}\big({\boldsymbol{x}^i_{c,3}}, {t^i_{c,3}}, \boldsymbol{w_3}\big)-\Tilde{u}^{(2)}\big({\boldsymbol{x}^i_{c,2}}, {t^i_{c,2}}, \boldsymbol{w_2}\big)\bigg)^2,\\
    \mathcal{L}_{fc,32}^{(1)}\big(\boldsymbol{w_3}, \boldsymbol{w_2}\big) =& \frac{1}{N_c} \sum_{i=1}^{N_c}\bigg(\Tilde{u}_x^{(3)}\big({\boldsymbol{x}^i_{c,3}}, {t^i_{c,3}}, \boldsymbol{w_3}\big)-\Tilde{u}_x^{(2)}\big({\boldsymbol{x}^i_{c,2}}, {t^i_{c,2}}, \boldsymbol{w_2}\big)\bigg)^2,\\[10pt]
    \mathcal{L}^{(3)}\big(\boldsymbol{w_3}, \boldsymbol{w_2}\big) =&\mathcal{L}_r^{(3)} + \mathcal{L}_0^{(3)} + \mathcal{L}_b^{(3)} + \mathcal{L}_{pc,32}^{(3)} + \mathcal{L}_{fc,32}^{(3)}
\end{align*}
For case 2.1, the well is controlled by bottom-hole pressure the results are displayed in Fig. \ref{fig:const_pres_2D}, and for case 2.2, with well being controlled by constant production rate the results are displayed in Fig. \ref{fig:const_pres_rates_pred_2D}. Fig. \ref{fig:rates_2D} shows the flow rates predictions for case 2.1 calculated using Darcy's equation \ref{darcy's} and the back-propagation to calculate $\frac{d\Tilde{p}}{dx}$.

\section{Conclusions and Future Work}
\label{sec:conclusions}
In this work we showed that physics-informed neural network (PINN) can approximate partial differential equations to very good accuracy both for pressure predictions and flow rates predictions. In relation to reservoir engineering, a PINN model can be regarded as a good replacement for the analytical solution and therefore enables us to perform instantaneous predictions. Utilizing the inverse features of PINNs and a production flow rates history data we managed to estimate reservoir properties such as permeability and boundary distance without having to shut in the well. Alternatively, such quantities are estimated using pressure transient or rate transient tests which both require shutting-in the well for number of days.

For the 2D case, we proposed a decomposition method which is essential to overcome the stiffness of the PDE. With that we were able to train a model that can accurately predict the solution of the diffusivity equation. Finally, it is imperative to mention that the inverse applications that were applied to the 1D case can be applied to the 2D case.

\section{Acknowledgements}
Part of this work was done during my internship at OXY in summer 2022. I would like to thank OXY for providing all the reservoir simulations support needed to achieve these results.

\bibliographystyle{unsrtnat}
\bibliography{references_zotero, references_extras}  

\end{document}